\documentclass[reprint,prl,twocolumn,superscriptaddress,showpacs,nofootinbib,hyperref,floatfix,preprintnumbers]{revtex4-1}
\usepackage{amssymb}
\usepackage{slashed}
\usepackage{amsmath}
\usepackage{braket}
\usepackage{float}
\usepackage[utf8]{inputenc}
\usepackage{graphicx}
\usepackage{subfigure}
\usepackage{latexsym}
\usepackage{epic}
\usepackage{eepic}
\usepackage{epsfig}
\usepackage{color}
\usepackage{slashed}
\usepackage{natbib}
\usepackage[colorlinks=true,citecolor=darkred,urlcolor=darkred, pdfborder={0 0 0}]{hyperref}
\usepackage[normalem]{ulem}

\definecolor{darkred}{rgb}{0.6,0,0}

\begin{document}
\begin{flushright}
\end{flushright}

\title{Signatures of $\tilde{R}_2$ class of Leptoquarks at the upcoming $ep$ colliders}

\author{Rojalin Padhan}
\email{rojalin.p@iopb.res.in}
\affiliation{Institute of Physics, Sachivalaya Marg, Bhubaneswar 751005, India}
\affiliation{Homi Bhabha National Institute, BARC Training School Complex,
Anushakti Nagar, Mumbai 400094, India}
\author{Sanjoy Mandal}
\email{smandal@iopb.res.in}
\affiliation{Institute of Physics, Sachivalaya Marg, Bhubaneswar 751005, India}
\affiliation{Homi Bhabha National Institute, BARC Training School Complex,
Anushakti Nagar, Mumbai 400094, India}
\author{Manimala Mitra}
\email{manimala@iopb.res.in}
\affiliation{Institute of Physics, Sachivalaya Marg, Bhubaneswar 751005, India}
\affiliation{Homi Bhabha National Institute, BARC Training School Complex,
Anushakti Nagar, Mumbai 400094, India}
\author{Nita Sinha}
\email{nita@imsc.res.in}
\affiliation{The Institute of Mathematical Sciences,
C.I.T Campus, Taramani, Chennai 600 113, India}
\affiliation{Homi Bhabha National Institute, BARC Training School Complex,
Anushakti Nagar, Mumbai 400094, India}
\preprint{\textbf{IP/BBSR/2019-11}}
\bibliographystyle{unsrt}

\begin{abstract}
We explore the signatures of the $\tilde{R}_2$ class of leptoquark~(LQ) models at the proposed $e^- p$ and $e^+p$ colliders. We carry out an analysis for the proposed colliders, LHeC and FCC-eh, with center of mass~(c.m.) energy 1.3 TeV and 3.46 TeV, respectively. For $\tilde{R}_2$ class of LQ models, there are a number of final states that can arise from LQs production and their subsequent decays. In this report we do a detailed cut-based analysis for the $l^{\pm}j$ final state. We also discuss the effect of polarized electron and positron beams on LQ production and in turn on $l^{\pm}j$ production. At LHeC,
the final state $l^+j$ has very good discovery prospect. We find that, with only 100 $\text{fb}^{-1}$ of data, one can probe LQs of masses upto 1.2 TeV with $5\sigma$ significance, even with a generic set of cuts.
On the contrary, at FCC-eh, one can probe LQ masses upto 2.2 TeV~(for $e^-$ beam) and 3 TeV~(for $e^+$ beam), at more than $5\sigma$ significance with luminosity $1000\,\text{fb}^{-1}$ and $500\,\text{fb}^{-1}$, respectively.
\end{abstract}

\maketitle

\section{Introduction} \label{introduction}
LQs are hypothetical particles which can emerge from the unification of quarks and leptons in the Pati-Salam model~\cite{Pati:1973uk}. LQs also exist in grand unified theories based on $SU(5)$~\cite{Georgi:1974sy} and $SO(10)$~\cite{Fritzsch:1974nn,Mohapatra:1979nn,Wilczek:1981iz}. TeV scale LQs can also exist in extended technicolor models~\cite{Dimopoulos:1979es,Dimopoulos:1979sp,Farhi:1980xs,Georgi:1981xw}. Under the Standard Model~(SM) representation, there are twelve types of LQs, six of them are scalar, while the other six are vector type of LQs~\cite{Dorsner:2016wpm}.
We consider the scalar LQ $\tilde{R}_2$ charged as (3,2,1/6) 
under SM gauge group. The advantage with $\tilde{R}_2$ type of scalar LQ is that in addition to the coupling with the lepton and jet, the model also has right handed~(RH) neutrinos coupled to the LQ. Hence, this model provides 
unique signatures, that can be tested in different collider and non-collider experiments.  Moreover, the $\tilde{R}_2$ also allows  matter stability at tree level~\cite{Arnold:2012sd,Cox:2016epl}. Discovery prospect of $\tilde{R}_2$ has been studied in dilepton final state at the LHC~\cite{Raj:2016aky}. There have been several searches for other types of LQs. In Ref.~\cite{Chandak:2019iwj}, both single and pair production of scalar LQs (with $Q_{EM}=1/3,\ 5/3$) have been studied with the decay mode, $LQ \to t l$. This  study shows that in the pair production channel, $5\sigma$ significance can be achieved at HL-LHC for LQ mass upto $1.7$ TeV. When LQ couples to only left-handed leptons, $5\sigma$ reach can extend beyond $2$ TeV. Pair production of LQs ($S_3$ and $U_1$) and their subsequent decay to  $t\bar{t} + \slashed{E}_T $  has been studied in Ref.~\cite{Vignaroli:2018lpq}. As stated in this search, a $U_1$ ($S_3$) LQ can be excluded with mass upto  1.96 TeV (1.54 TeV) or observed with mass upto 1.83 TeV (1.41 TeV), at HL-LHC. In Ref.~\cite{Allanach:2019zfr}, pair production LQs ($S_3$) and their decay to a $\mu\mu jj$ final state has been explored at future hadron colliders. The $5\sigma$ discovery sensitivity  of HL-LHC have been estimated to be  1.9 TeV.	In Ref.~\cite{Mandal:2018kau}, a recast of heavy resonance search at LHC has been performed in the $\tau\tau$ and $\tau\nu$ channels. Also, the exclusion limit on the coupling of a scalar LQ ($S_1$, which is motivated by the B-anomalies) as a function of its mass, has been placed. A review on other class of LQs, which are proposed to address B-anomalies has been presented in Ref.~\cite{Cerri:2018ypt}.

LQ's can be most easily tested at $ep$ colliders. At $ep$ colliders such as LHeC~\cite{AbelleiraFernandez:2012cc,Azuelos:2018syu} and FCC-eh~\cite{Zimmermann:2014qxa,Azuelos:2018syu}, LQs can be resonantly produced. LHeC and FCC-eh are the proposed $ep$ colliders, planned to operate with c.m. energies $\sqrt{s}=1.3$~TeV and $\sqrt{s}=3.46$~TeV, respectively. LHeC~(FCC-eh) will use electron and possibly positron beams of 60 GeV,
to collide with the 7 TeV~(50 TeV) proton beam. There are  numerous  important phenomenological beyond the standard model studies for $ep$ colliders which have been performed and listed in Ref.~\cite{LHeCFCCehBSM}. 

For the specific type of LQ model, that we consider in this paper, the LQ can decay to a lepton and a jet, as well as, to a jet and a RH neutrino. 
The decay of the LQ into a lepton and a jet, and the decay to a jet and RH neutrino with subsequent decays of RH neutrino  give rise to many possible final states. In this work, we have studied in detail the final state $l^{\pm}j$. We have considered both the scenarios of electron and positron beams.
We show that with judicious application of selection cuts, the final state $l^{\pm}j$ has very good discovery prospects at LHeC and FCC-eh colliders. We find that, we can probe the LQ mass up to 1.2 TeV at more than $5\sigma$ significance with $e^+$ beam at LHeC. For FCC-eh, we can easily probe LQ mass up to 2.2 TeV  (3 TeV) with  $e^-$ ($e^+$) beams.

The paper is organized as follows: First we review the model and the existing constraints on LQs. Following this, we discuss the production of a LQ at an $ep$ collider and compare with that at the LHC. In the subsequent sections, we present a detailed collider analysis and discuss the discovery prospects of the final state $l^{\pm}j$. Finally, we conclude.
\section{Model}
\label{model}
We consider an extension of the SM with a single scalar LQ, $\tilde{R_2}(3, 2, 1/6) = (\tilde{R_2}^{\frac{2}{3}},\tilde{R_2}^{-\frac{1}{3}})^T$. This is a genuine LQ ($F=3B+L=0$). The superscript of $\tilde{R_2}$ denotes the electromagnetic~(em) charge. In the presence of the RH neutrinos $N_R$~(1,1,0), the 
LQ has additional interaction \cite{Dorsner:2016wpm,Buchmuller:1986zs,Buchmuller:1986iq,Becirevic:2016yqi},
\begin{align}
 \mathcal{L}=-Y_{ij}\bar{d}_{R}^{i}\tilde{R}_{2}^{a}\epsilon^{ab}L_{L}^{j,b}+
 Z_{ij}\bar{Q}_{L}^{i,a}\tilde{R}_{2}^{a}N_{R}^{j}+h.c., 
 \label{eq1}
\end{align}
where $a,b=1,2$ are $SU(2)_L$ indices.
Upon expansion, the Lagrangian becomes
\begin{align}
 &\mathcal{L}=-Y_{ij}\bar{d}_{R}^{i}e_{L}^{j}\tilde{R}_{2}^{2/3}+(YU_{\text{PMNS}})_{ij}\bar{d}_{R}^{i}\nu_{L}^{j}
 \tilde{R}_{2}^{-1/3}+\\ \nonumber
 & (V_{\text{CKM}}Z)_{ij}\bar{u}_{L}^{i}N_{R}^{j}\tilde{R}_{2}^{2/3}+Z_{ij}\bar{d}_{L}^{i}
 N_{R}^{j}\tilde{R}_{2}^{-1/3}+h.c.,
 \label{eq2}
\end{align}
where $i,j=1,2,3$ are flavor indices. $Y$ and $Z$ are the Yukawa couplings. $U_{\text{PMNS}}$ and $V_{\text{CKM}}$ represent the Pontecorvo-Maki-Nakagawa-Sakata and the Cabibbo-Kobayashi-Maskawa matrices. For simplicity, we assume that both the Yukawa couplings are diagonal, $Y_{ij}=\delta_{ij}Y_{ii}$ and $Z_{ij}=\delta_{ij}Z_{ij}$, where $i,\,j=1,2,3$.
Hence in our model a LQ couples to the same generation of lepton and quark. Although, most of the collider bounds on LQ mass and coupling are derived assuming that only one generation is  present at a time, our model can have non-zero couplings of LQ to fermions of more than one generation. In the next section, we review the existing constraints on LQ mass and couplings.
\section{Constraints on LQ mass and couplings}
\label{sec:constraints}
Tight constraints exist on $\tilde{R}_2$ type of scalar LQ's mass and coupling from both the collider experiments as well as low energy experiments such as those searching for atomic parity violation and lepton flavour violating decays, $K_L\to\mu^{-}e^+$ etc. Below, we summarize present bounds on LQ mass and couplings.
\subsection{Atomic Parity Violation}
There are tight bounds on Yukawa couplings $Y_{de}$ and $Y_{ue}$ from atomic parity violation~(APV) experiments, which  have been explained in Ref.~\cite{Dorsner:2014axa,Babu:2019mfe}. Following Ref.~\cite{Dorsner:2014axa}, the present bound on the coupling $Y$ from atomic parity violation are $Y_{de}<0.34\frac{M_{\text{LQ}}}{1\,\text{TeV}}$ and $Y_{ue}<0.36\frac{M_{\text{LQ}}}{1\,\text{TeV}}$. Hence for larger LQ mass the allowed values of Yukawa couplings will also be large. Note that, these bounds are derived with the assumption that only one of the Yukawa coupling is present at a time.
\subsection{$K_{L}\to\mu^{-}e^{+}$}
The tree level lepton flavour violating~(LFV) process $K_L\to\mu^-e^+$ gives tight constraints on the diagonal couplings of $\tilde{R}_2^{\frac{2}{3}}$. Specifically this tree level process constrains the product of Yukawa couplings $|Y_{s\mu}Y_{de}^*|$. Following Refs.~\cite{Dorsner:2014axa,Dorsner:2011ai}, the experimental upper bound on the decay mode $K_{L}\to\mu^-e^+$ results in the bound on the product of Yukawas and is given by $|Y_{s\mu}Y_{de}^*|\leq 2.1\times 10^{-5}\frac{M_{\text{LQ}}}{1\,\text{TeV}}$. Therefore, couplings of first two generations are tightly constrained. \\
\subsection{Collider bounds}
 The present tightest collider bounds on LQs come from LHC~\cite{Sirunyan:2017yrk,Aaboud:2016qeg,Sirunyan:2018kzh,Sirunyan:2018jdk,ATLAS:2017vke,Sirunyan:2018btu}. LHC has specifically looked for the final states $pp\to \text{LQ} \,\text{LQ}\to \ell j\ell j$ with the assumption that LQ  decays to the final state $\ell j$ with $100\%$ branching ratio. LHC  constrains first~\cite{Sirunyan:2018btu}, second~\cite{Aaboud:2016qeg} and third generation~\cite{Sirunyan:2017yrk} of LQs considering $\ell$ to be $e,\mu$ or $\tau$. From the non-observation of any new physics at LHC, the LQs with mass up to 1.435 TeV  have been ruled out for the first generation at $95\%$ C.L~\cite{Sirunyan:2018btu}.
 \begin{figure}
\centering
\includegraphics[width=0.3\textwidth]{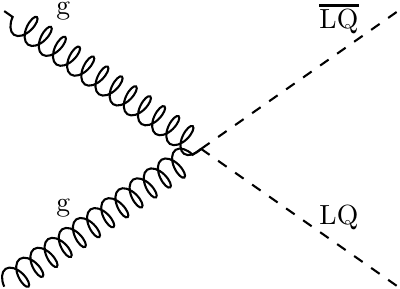}
\hspace{1cm}
\includegraphics[width=0.28\textwidth]{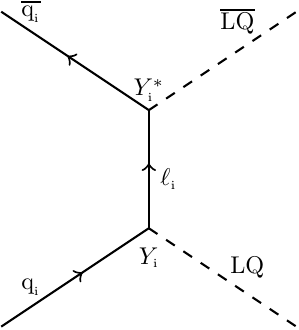}
\caption{\small{Upper panel: Feynman diagram for the gluon-initiated LQ pair-production process at LHC. Lower panel: the same, but for the quark-initiated processes.}}
\label{feynman diagram for LHC}
\end{figure}

\begin{figure}
\centering
\includegraphics[width=0.5\textwidth]{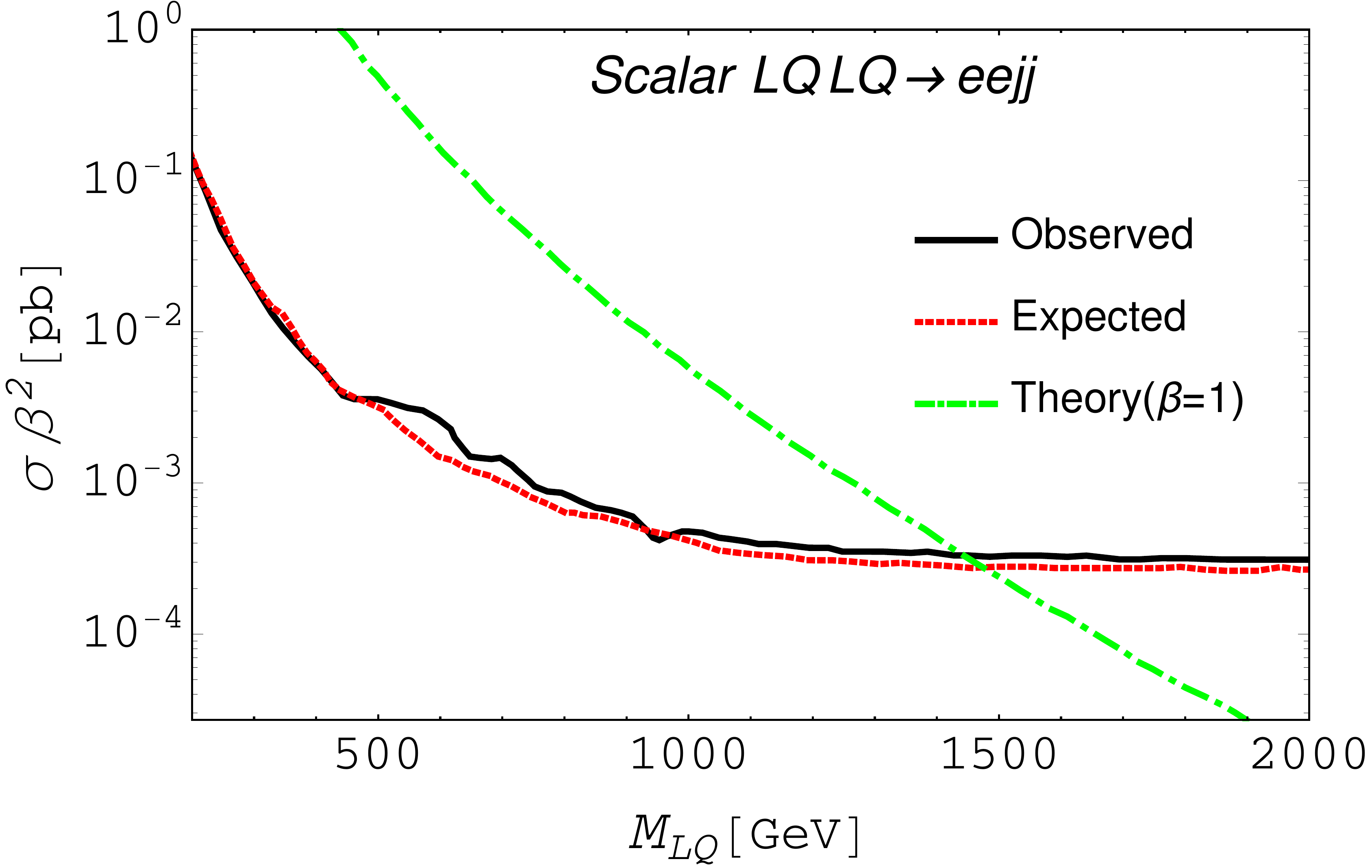}
\caption{\small{Present limit on scalar LQ pair-production times branching fraction to eejj final state as a function of mass for first generation LQs from CMS experiment with $\sqrt{s}=13$~TeV and luminosity $35.9\,\text{fb}^{-1}$. The red and black lines are the expected and observed limits. The green line is the theory prediction.}}
	\label{Present Limit on LQ}
\end{figure}

In Fig.~\ref{feynman diagram for LHC}, we have shown the Feynman diagrams for LQ pair production at LHC. For smaller values of Yukawa coupling $Y$, LQ pair production is dominated by the gluon-initiated diagrams, whereas for relatively larger Yukawa coupling $Y$, the t-channel quark-initiated diagram can dominate.
In Fig.~\ref{Present Limit on LQ}, we have shown present limit on scalar LQ pair-production times branching fraction to $ej$ final state as a function of LQ mass. This bound corresponds to that from CMS experiment with $\sqrt{s}=13$~TeV and integrated luminosity $35.9\,\text{fb}^{-1}$~\cite{Sirunyan:2018btu}. Black and red lines represent observed and expected limits. Note that, Ref.~\cite{Sirunyan:2018btu} also studies the $e\nu j j$ channel.
For $\beta = 0.5$, using the $e\nu jj$ channel alone, LQ masses are excluded below
1.195 TeV. It also gives a limit on LQ mass combining both the channels $eejj$ and $e\nu j j$. In our analysis we have chosen the values of Yukawa couplings and LQ mass according to this combined limit presented in Fig.~9 of Ref.~\cite{Sirunyan:2018btu}.

\begin{table}[b]
\scalebox{1}{
\begin{tabular}{|c|c|c|c|c|}
\hline Benchmarks&$M_{\text{LQ}}$&$M_{N_1}$& $Y$& $Z$\\ \hline
$\text{BP}_1$&1000&$ 100 $& $(0.34,0,0)$& $(1.03,0,0)$\\ \hline
\end{tabular}}
\caption{Values of different model parameters used in the analysis for $M_{\text{LQ}}=1$~TeV.}
\label{table:BP1}
\end{table}
Note that, assuming $100\%$ branching ratio to $ej$ final state~(which implies $Z_{11}=0$), bound on LQ mass is 1.435 TeV. If the branching ratio $\beta$ to $ej$ is less than $100\%$~(which is of course possible if LQ has additional interactions such as LQ-$\nu j$, LQ-$jN_i$), the bound on LQ mass can be lowered. For example, to be consistent with LHC, flavour and APV constraints, for LQ mass of 1 TeV, one can choose the coupling $Y_{11}=0.34$ and $Z_{11}=1.03$. We have shown this benchmark $\text{BP}_1$ in Table~\ref{table:BP1}. For different LQ masses we have chosen different benchmark points consistent with all the existing constraints, which  have been summarized in Table~\ref{table:mlq_y_z}.
\begin{table}[h]
	\scalebox{1.3}{ \begin{tabular}{|c|c|c|}
			\hline
	$M_{LQ}$&$Y_{11}$&$Z_{11}$	\\	\hline  
	687&0.233&1.29\\ \hline              
	860&0.29&1.27\\ \hline
	1000&0.34&1.03\\ \hline
	1110&0.377&0.84\\ \hline
	1204&0.41&0.65\\ \hline
	\end{tabular}}
	\caption{LQ mass, corresponding to the maximum allowed value of $Y_{11}$ consistent with APV and  lower bound on $Z_{11}$ according to LHC constraint. } \label{table:mlq_y_z}	
\end{table}

In the next section, we discuss about the LQ production at colliders.
\section{LQ production}
\label{LQ production}
\begin{figure}
\centering
\includegraphics[width=0.35\textwidth]{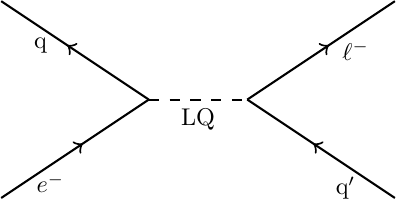}
\includegraphics[width=0.3\textwidth]{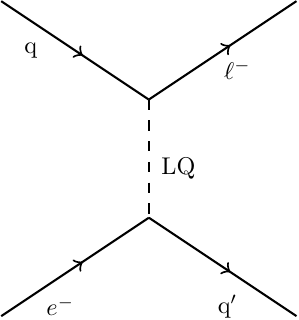}
\caption{\small{Feynman diagram for the $\ell j$ production in $e^-p$ collider.}}
\label{feynman diagram at LHeC for process lj}
\end{figure}
\begin{figure}
\centering
\includegraphics[width=0.35\textwidth]{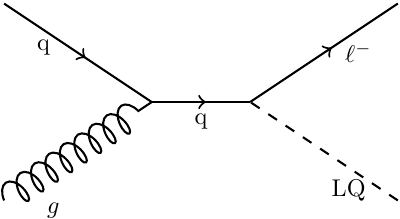}
\hspace{0.5cm}
\includegraphics[width=0.3\textwidth]{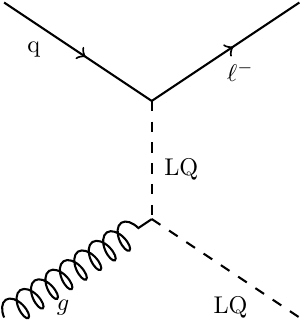}
\caption{\small{Single LQ production at LHC}.}
\label{Single LQ production at LHC}
\end{figure}
\begin{figure}
\centering	\includegraphics[width=0.45\textwidth]{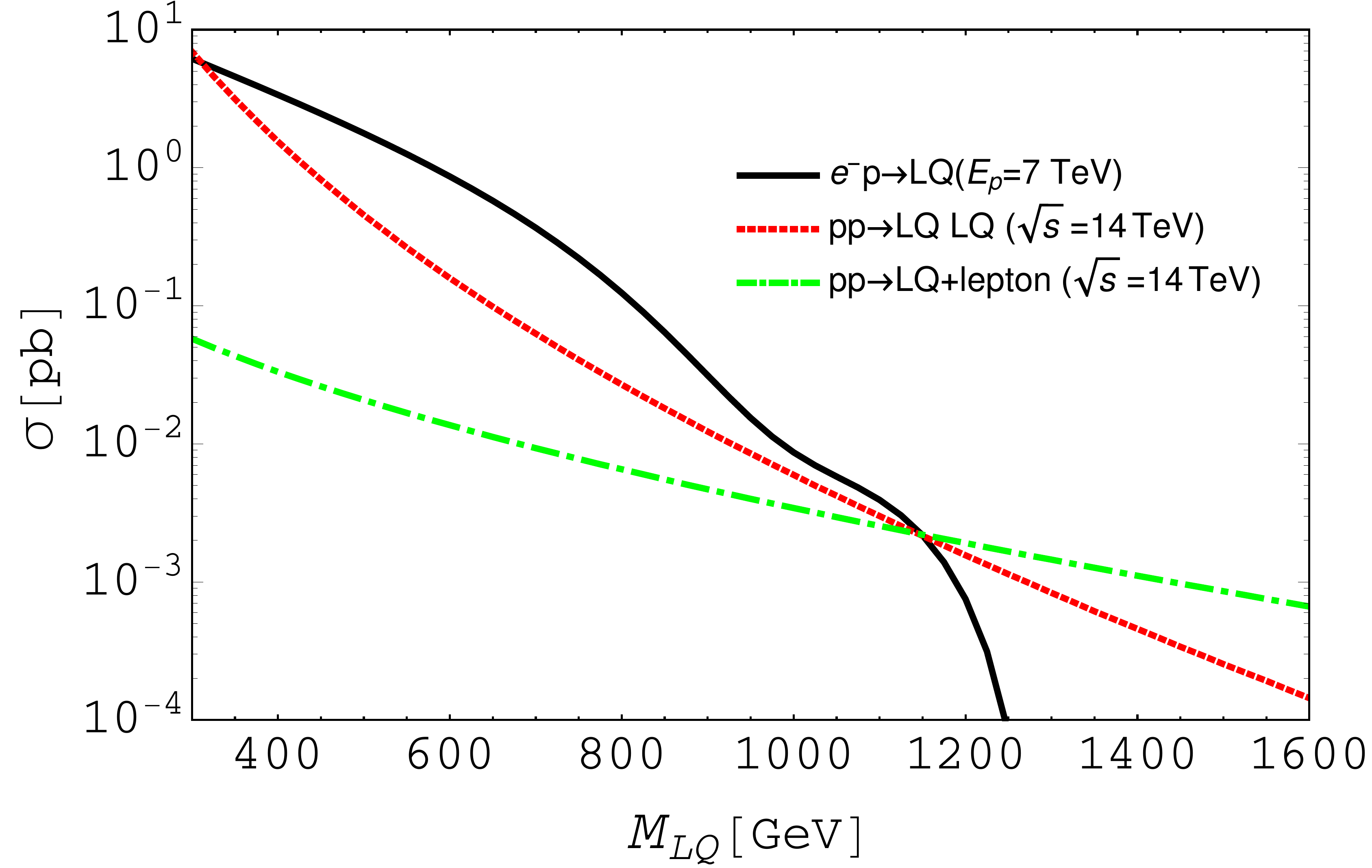}
\includegraphics[width=0.45\textwidth]{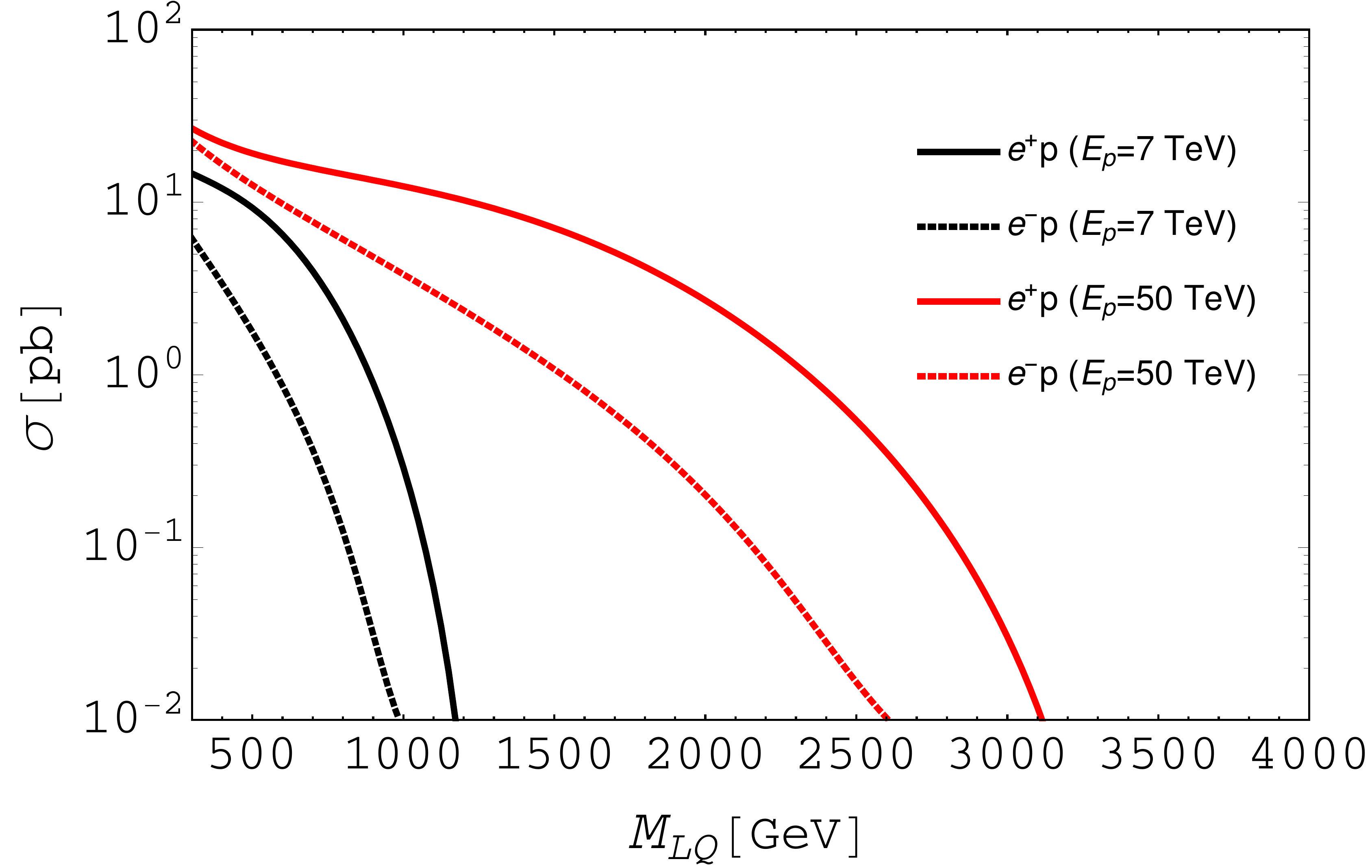}
\caption{\small{Upper Panel: comparison of single LQ production at LHeC with respect to single or pair production of LQ at LHC. Lower Panel: cross section for single LQ production at $e^+p$ and $e^{-}p$ collider for various proton beam energies. The electron or positron beam has been fixed at 60 GeV. In both these cases, the coupling $Y_{de}$ has been varied as $0.34\frac{M_{\text{LQ}}}{1\,\text{TeV}}$, in agreement with the APV constraints. For these plots we have considered $Z_{11}=0$.}}
	\label{Cross section at LHC and LHeC}
\end{figure}

\begin{figure}
\centering
\includegraphics[width=0.45\textwidth]{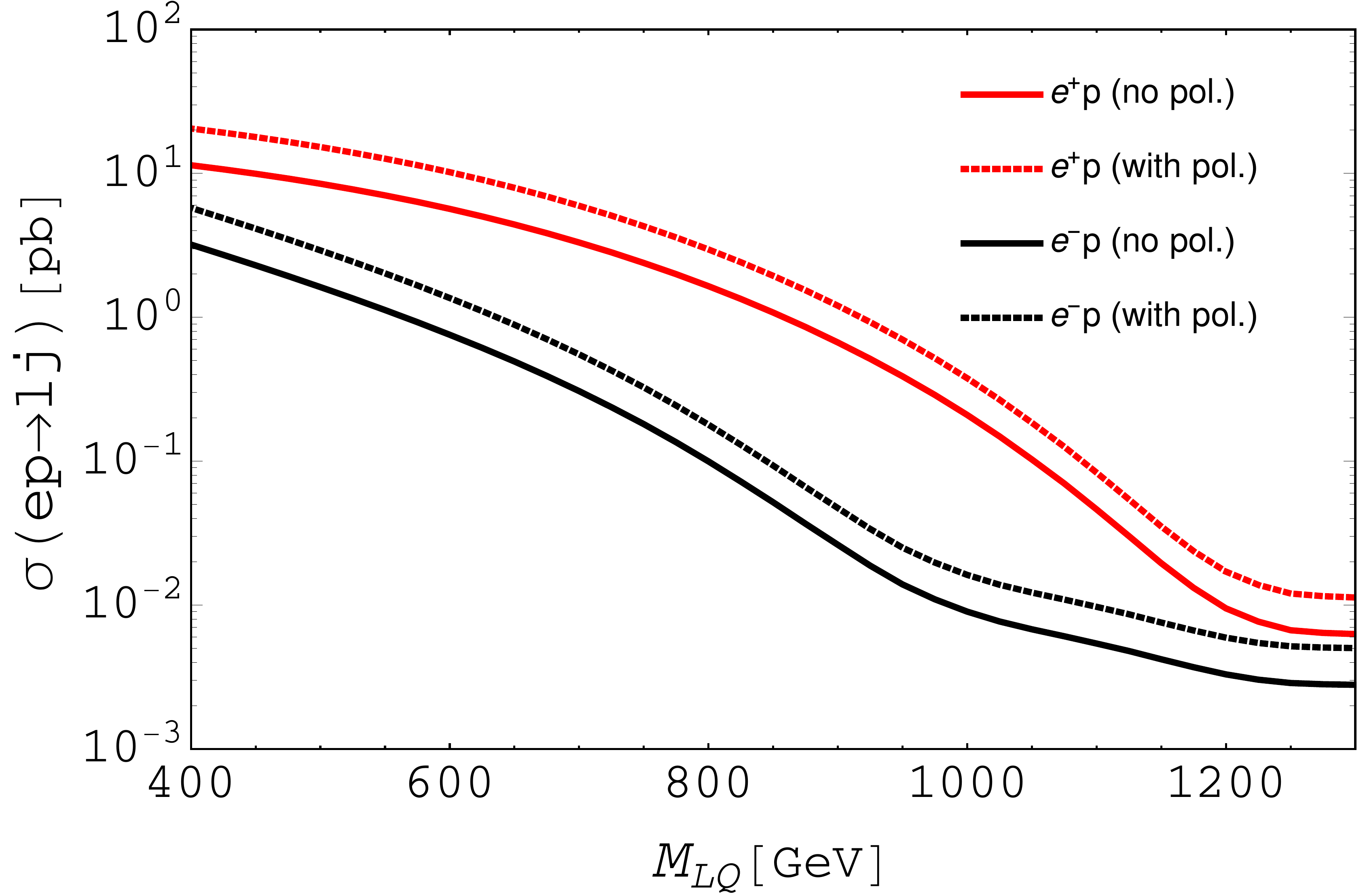}
\includegraphics[width=0.45\textwidth]{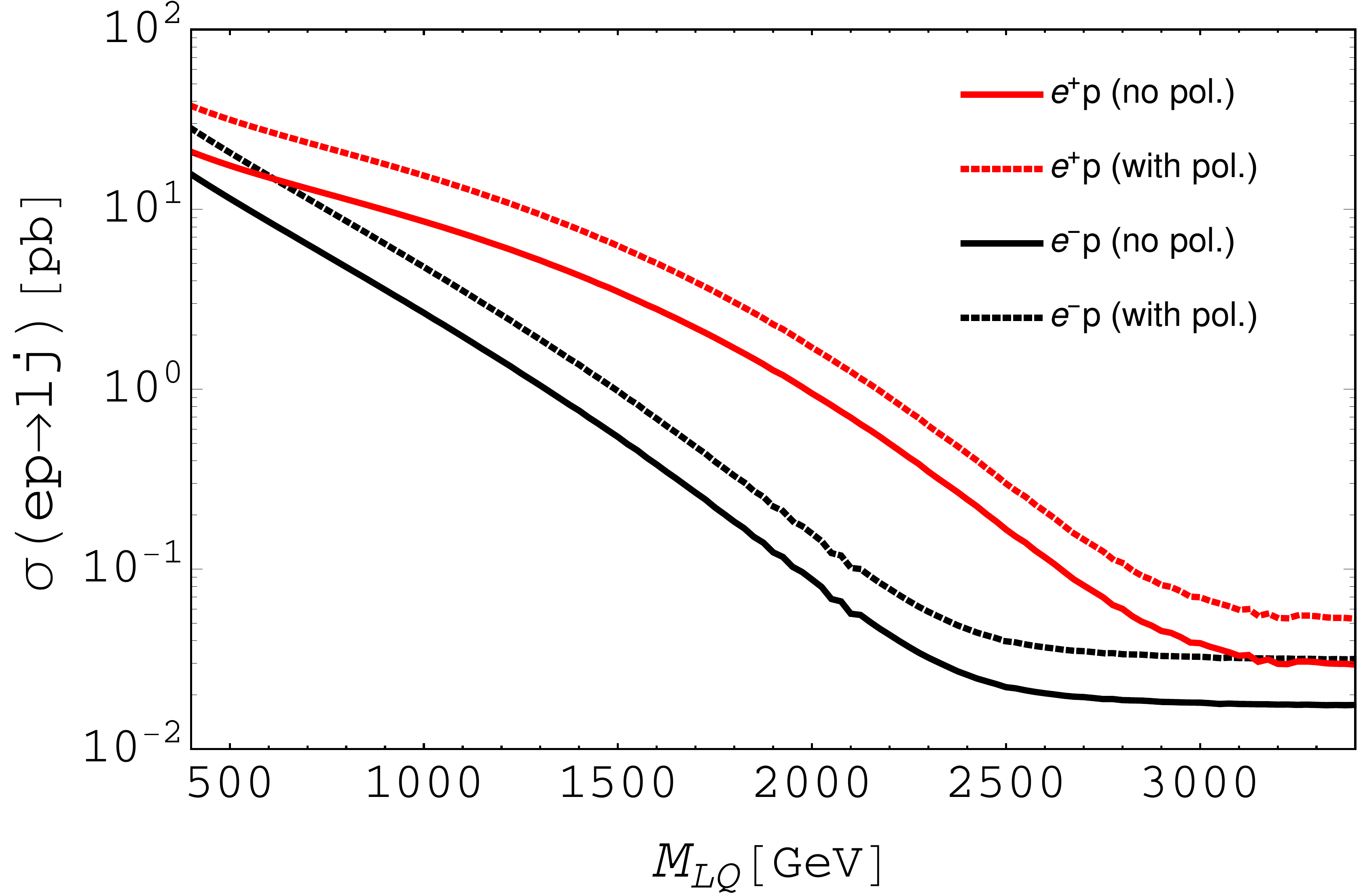}
\caption{\small{Production cross-section of $l^-j$~($l^+j$) with and without $80\%$ left(right)- polarized electron~(positron) beam.  The dotted and solid lines represent the variation of production cross-section with and without the polarized electron or positron beam. Upper panel is for LHeC and lower panel is for FCC-eh. In both these cases, the coupling $Y_{de}$ has been varied as $0.34\frac{M_{\text{LQ}}}{1\,\text{TeV}}$, in agreement with the APV constraints and we have fixed $Z_{11}=0$.}.}
\label{polarization effect}
\end{figure}
\begin{figure}
	\includegraphics[width=8.2cm,height=6.5cm]{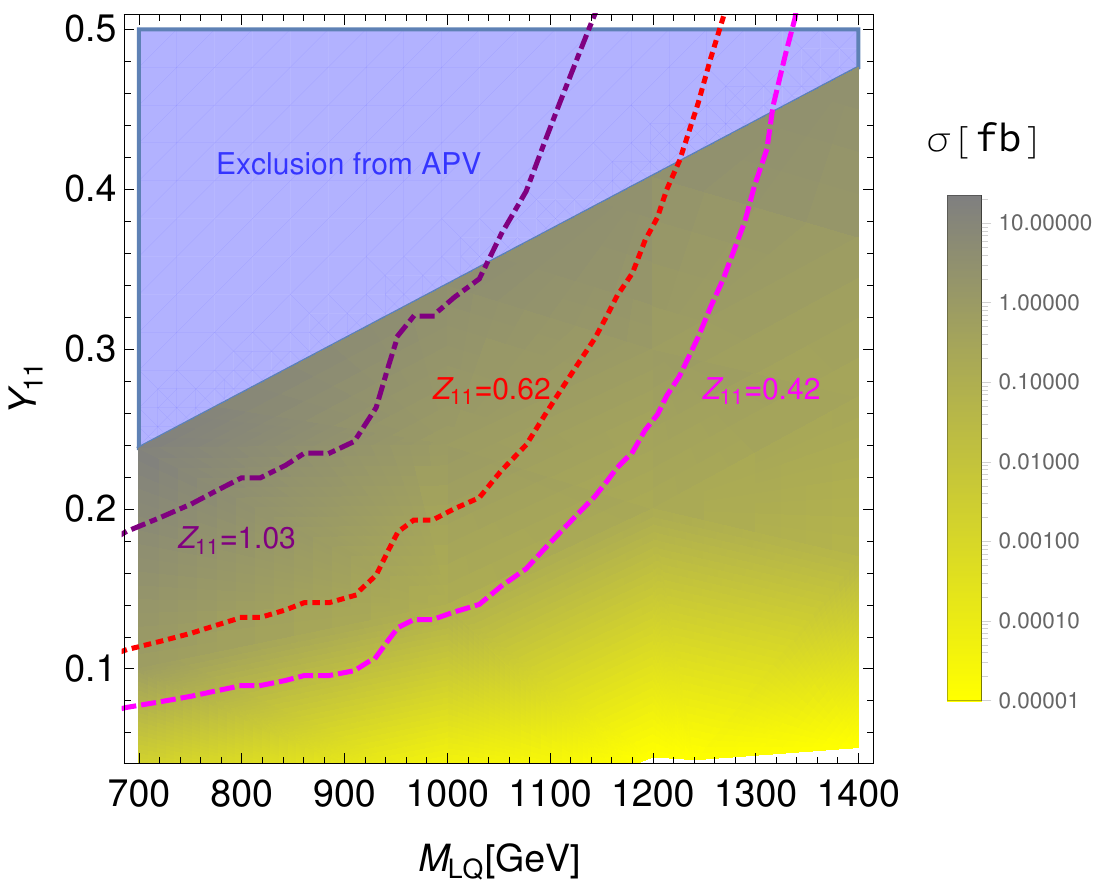}
	\caption{Variation of $\sigma(e^- p \to l^-  j)$ in $Y_{11}-M_{\text{LQ}}$ plane. Blue region is disallowed from atomic parity violation. The upper region  corresponding to each of the dashed~($Z_{11}=0.42$), dotted~($Z_{11}=0.62$), and dotted-dashed~($Z_{11}=1.03$) line 
		is disallowed from 13 TeV LHC search~\cite{Sirunyan:2018btu}, where we consider the combined limit on branching ratio from \cite{Sirunyan:2018btu}. The colour bar indicates the cross-section for $e^- p \to l^- j$ for  60 GeV electron and 7 TeV proton beam. We consider $\text{BP}_1$ for this plot. }  \label{fig:mycx}
\end{figure}
At an $ep$ collider, scalar LQs can be resonantly produced through s-channel process,  as shown in the upper panel of Fig.~\ref{feynman diagram at LHeC for process lj}. LQs can also  mediate t-channel process $ep \to l j$, as shown in the lower panel of Fig.~\ref{feynman diagram at LHeC for process lj}. Single or pair production of LQs is also possible at $pp$ colliders such as LHC. We have shown the Feynman diagrams for pair production and single production of LQs at LHC in Fig.~\ref{feynman diagram for LHC} and Fig.~\ref{Single LQ production at LHC}, respectively. A comparison between production cross-section of LQ at LHeC and LHC has been presented in the upper panel of Fig.~\ref{Cross section at LHC and LHeC}. Here we have shown the  cross-section for single LQ production at LHeC, as well as  both the single and
	pair production cross-section at LHC, for various LQ masses. For this comparison, electron and proton beams are fixed at 60 GeV and 7 TeV, respectively. For LQ mass upto 1.2 TeV, the single LQ production at $ep$ collider clearly dominates over the single and pair production of LQ at LHC. For LQ mass $M_{\text{LQ}}>1.2$~TeV, single LQ production at LHC dominates. In the lower panel of Fig.~\ref{Cross section at LHC and LHeC}, we have given the comparison of single LQ production at $e^-p$ and $e^+p$ colliders. For this comparison we have fixed the electron or positron beam at 60 GeV and for proton beam we have taken 7 TeV and 50 TeV. From this plot, it is evident   that LQ production at $e^+p$ collider is  larger than that at the $e^-p$ collider for the chosen mass range. This occurs as the $\frac{2}{3}$ charged LQ couples with $e^--\bar{d}$ and $e^+-d$, respectively and a quark parton distribution function is larger than that for anti-quark.

We also compute the production cross-section for the channel $lj$ with both polarized and unpolarized electron and positron beams to show how much the cross-sections differ. There is a relative enhancement in $l^-j$~($l^+j$) production cross-sections at $e^-p$~($e^+p$) collider when the electron~(positron) beam is dominantly left~(right)-polarized. We have shown our results in Fig.~\ref{polarization effect} for both LHeC and FCC-eh case. We can see that the production rates improve by almost a factor of 2 over the entire range of the LQ mass in the case of polarized electron or positron beams. This enhancement occurs due to the couplings $e^-_L-\tilde{R}_2^{\frac{2}{3}}-\bar{d}$ and $e^+_R-\tilde{R}_2^{-\frac{2}{3}}-d$ at $e^-p$ and $e^+p$ colliders, respectively. Hence at $e^-p$ and $e^+p$ colliders, LQ predominantly couples with left polarized electron and right polarized positron, respectively.

In Fig.~\ref{fig:mycx}, we present a density plot which shows the variation of cross-section for the process  $e^- p \to l^-  j$ with the variation of $\tilde{R_2}-e-d$ coupling ($Y_{11}$) and mass of the LQ ($M_{\text{LQ}}$). Here the $\tilde{R_2}$-$\text{N}_R-u$ coupling ($Z_{11}$) is fixed to 1.03 and mass of the right-handed neutrino ($M_{N_1}$) is assumed to be 100 GeV. The dashed curves represent the constraints on $Y_{11}$, for  $Z_{11}$ equal to 0.42, 0.62 and 1.03, from LHC search~\cite{Sirunyan:2018btu}. Region below each dashed curve is allowed for respective value of $Z_{11}$. Blue region in $Y_{11}-M_{\text{LQ}}$ plane is disallowed due to limit on $Y_{11}$ from APV. Here we consider c.m. energy $\sqrt{s}=1.3$~TeV. As we can see, a 700 GeV mass LQ is still allowed, for $Y_{11}=0.238$ and $Z_{11}=1.03$. Finally, in the next section we present our analysis for the model signature $ep \to l j$. 
\section{Analysis}
\label{analysis}
We consider the process $e^{\pm} p \to l^{\pm}  j$ as the signal. The dominant background for this process comes from the SM processes $e^{\pm} p \to l^{\pm}  j ,\ l^{\pm}  jj$. The analysis is carried out with $\sqrt{s}=1.3$~TeV of LHeC and $\sqrt{s}=3.46$~TeV of FCC-eh.
To simulate the signal samples, we implement the model in FeynRules(v2.3)~\cite{Alloul:2013bka}. The UFO output is then fed in \texttt{MadGraph5\_aMC@NLO}(v2.6)~\cite{Alwall:2014hca} that generates the parton-level events. We perform parton showering and hadronization with Pythia6~\cite{Sjostrand:2006za} and carry out the detector simulations with Delphes(v3.4.1)~\cite{deFavereau:2013fsa}. Finally data analysis and plotting is done in ROOT(v6.14/04)~\cite{Brun:1997pa}.

We apply the following basic cuts $P_{T}(l) > 10$ GeV, $P_T(j) > 20$ GeV, $|\eta (l)| < 5$,  $|\eta (j)| < 5$, $\Delta R_{ll} > 0.4$, $\Delta R_{jj} > 0.4$, $\Delta R_{lj} > 0.4$ for event generation in MadGraph5. 

\begin{figure}
 	\centering	
 	\includegraphics[width=0.45\textwidth,height=0.3\textheight]{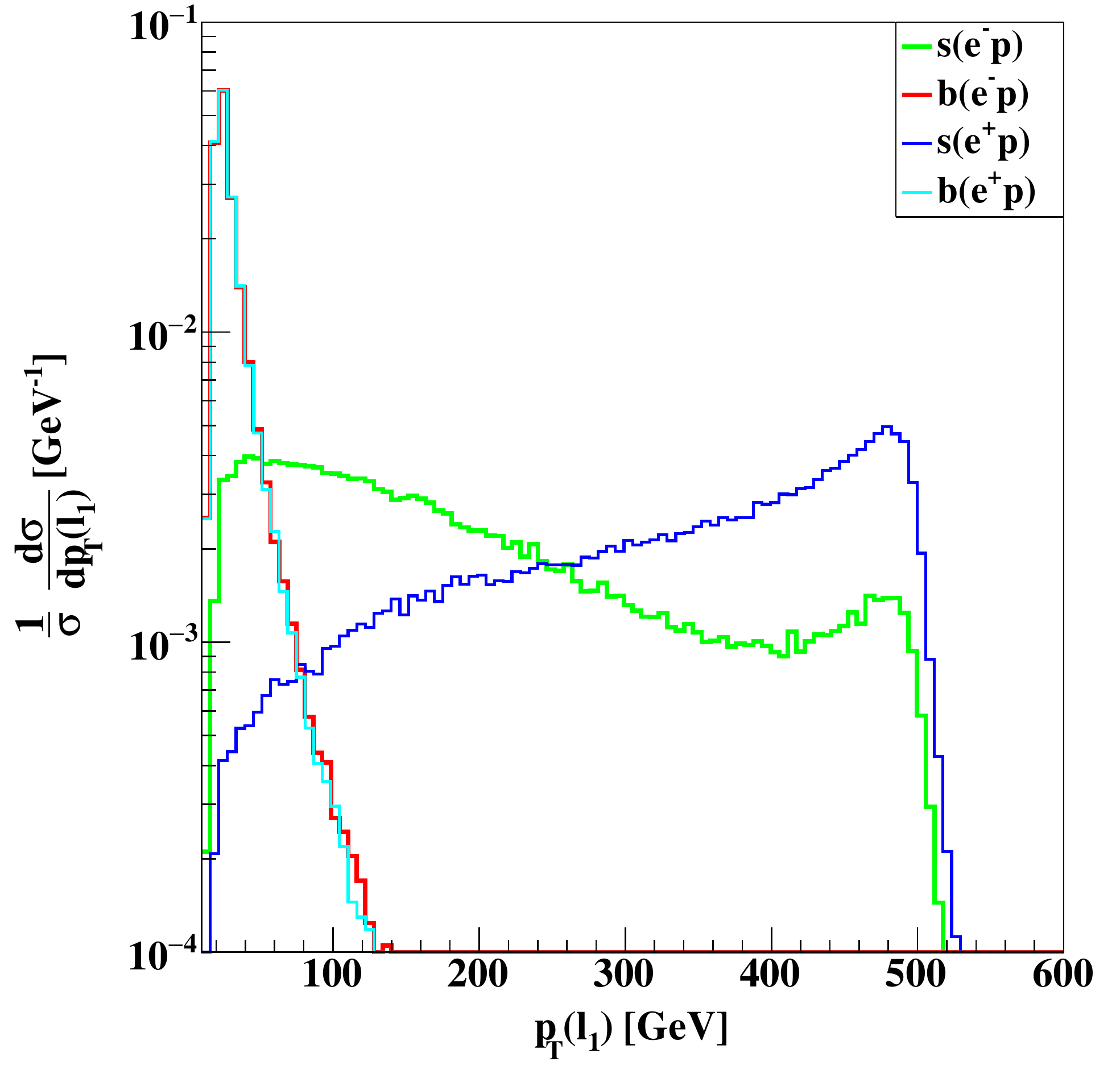}
 	\includegraphics[width=0.45\textwidth,height=0.3\textheight]{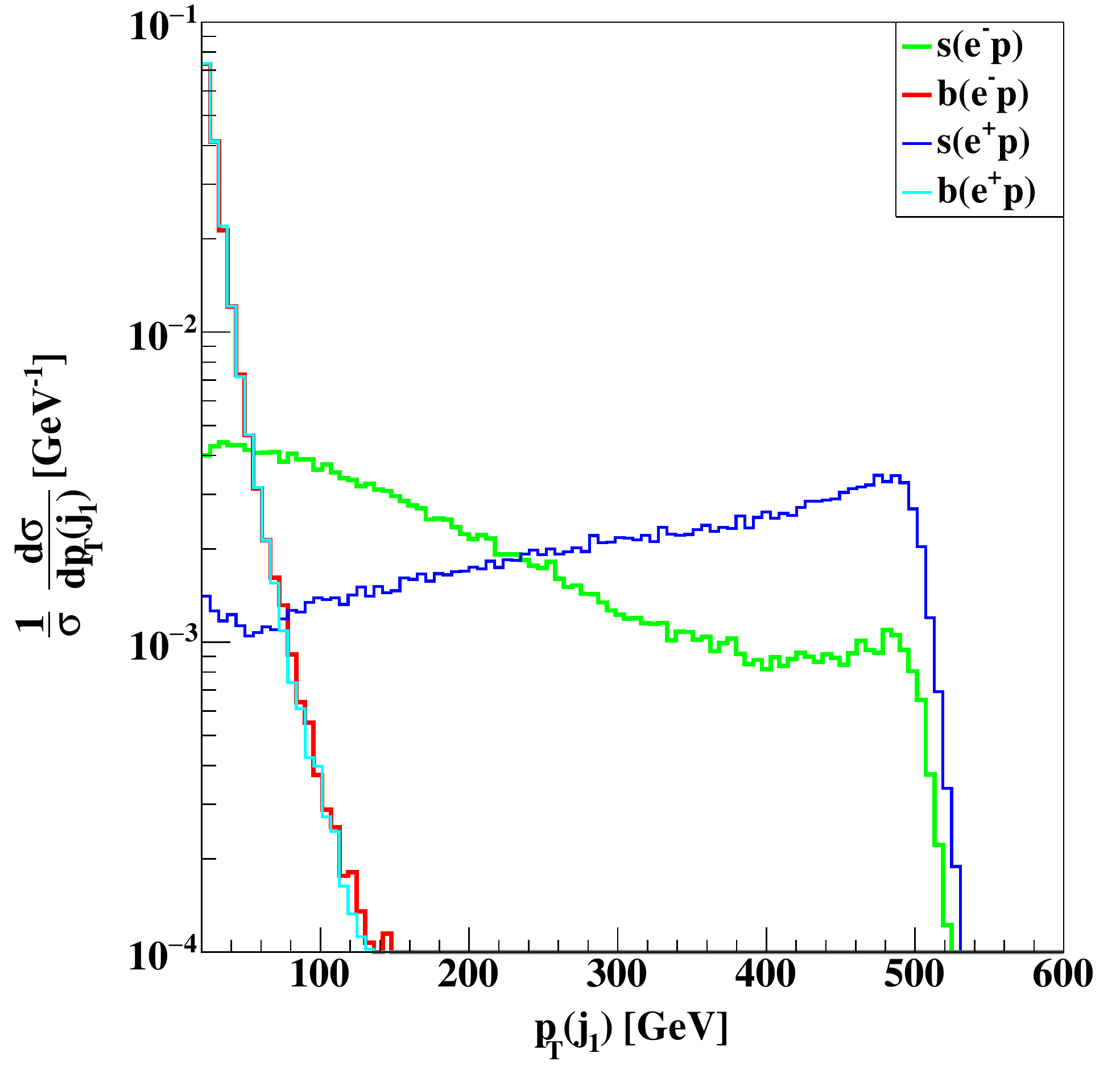}
 	\includegraphics[width=0.45\textwidth,height=0.3\textheight]{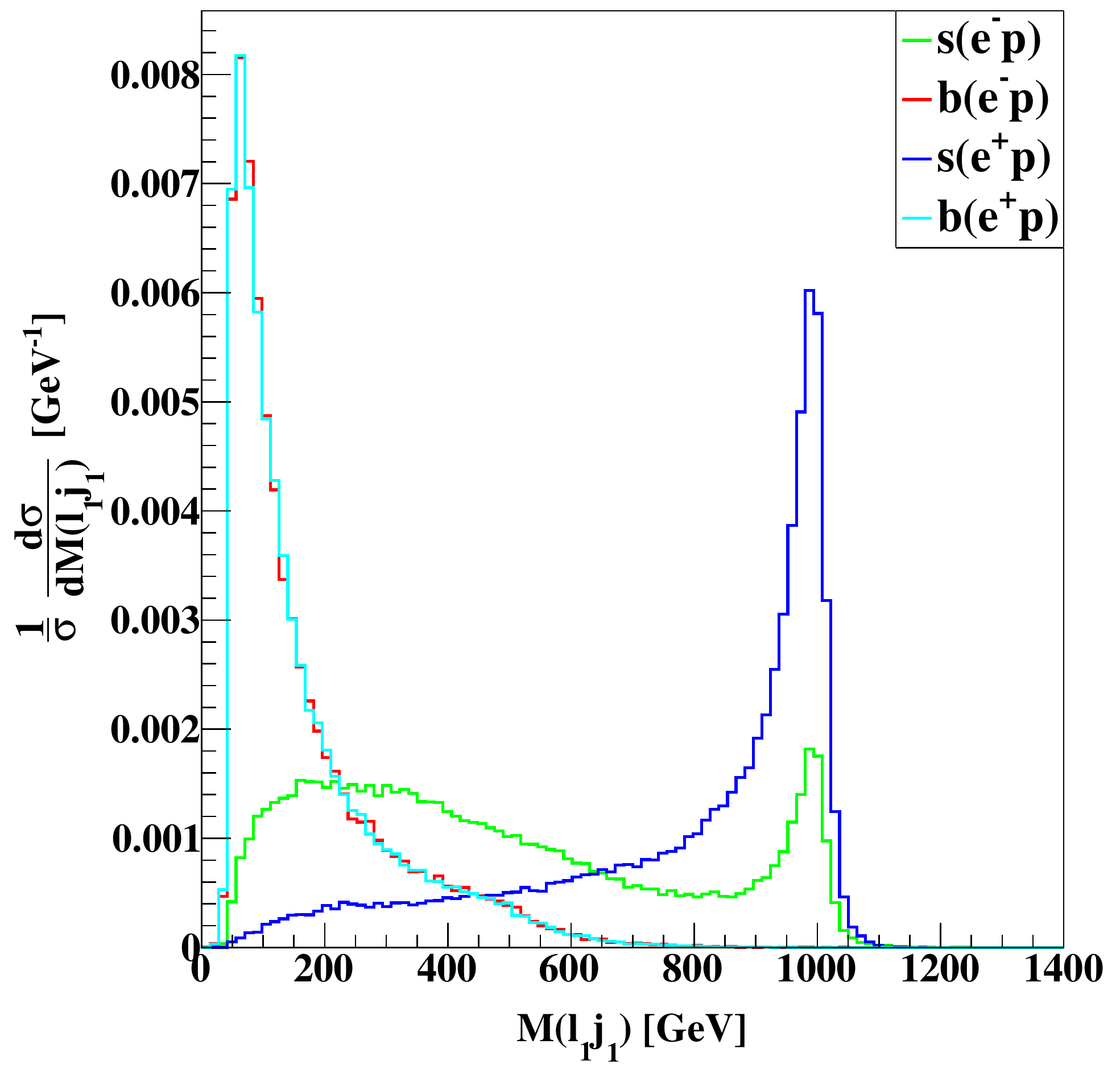}
 	\caption{\small{Normalised distribution of transverse momentum of leading lepton $p_T(l_1)$, leading jet $p_T(j_1)$, invariant mass of leading lepton and jet $M(l_1j_1)$ for both signal and background with c.m.energy $\sqrt{s} = 1.3$ TeV. Here the parameter set given in Table.~\ref{table:BP1} is considered.}}
 	\label{fig:distribution_lhec}
 \end{figure}
 \begin{figure}
 	\centering	
 	\includegraphics[width=0.45\textwidth,height=0.3\textheight]{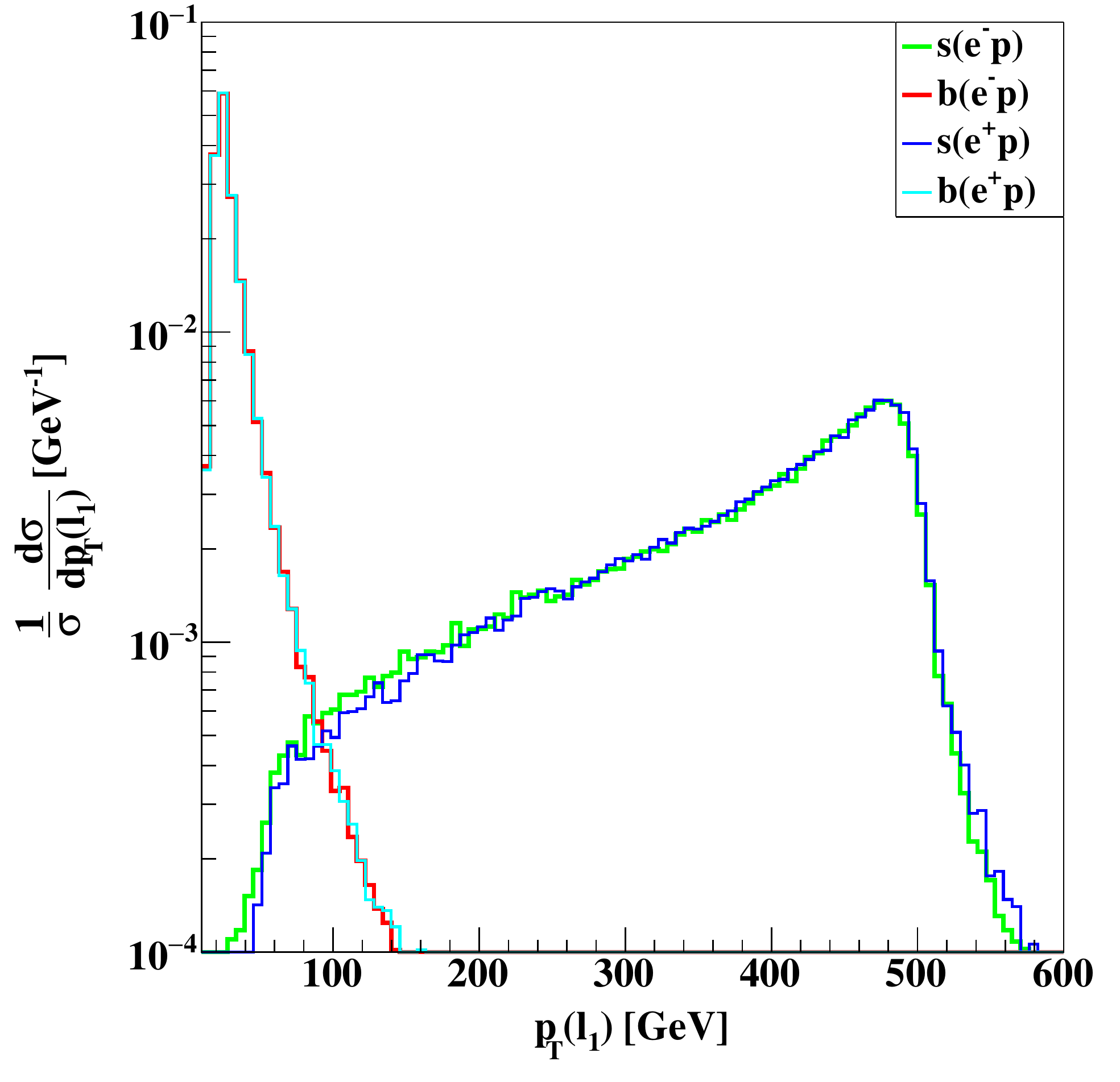}
 	\includegraphics[width=0.45\textwidth,height=0.3\textheight]{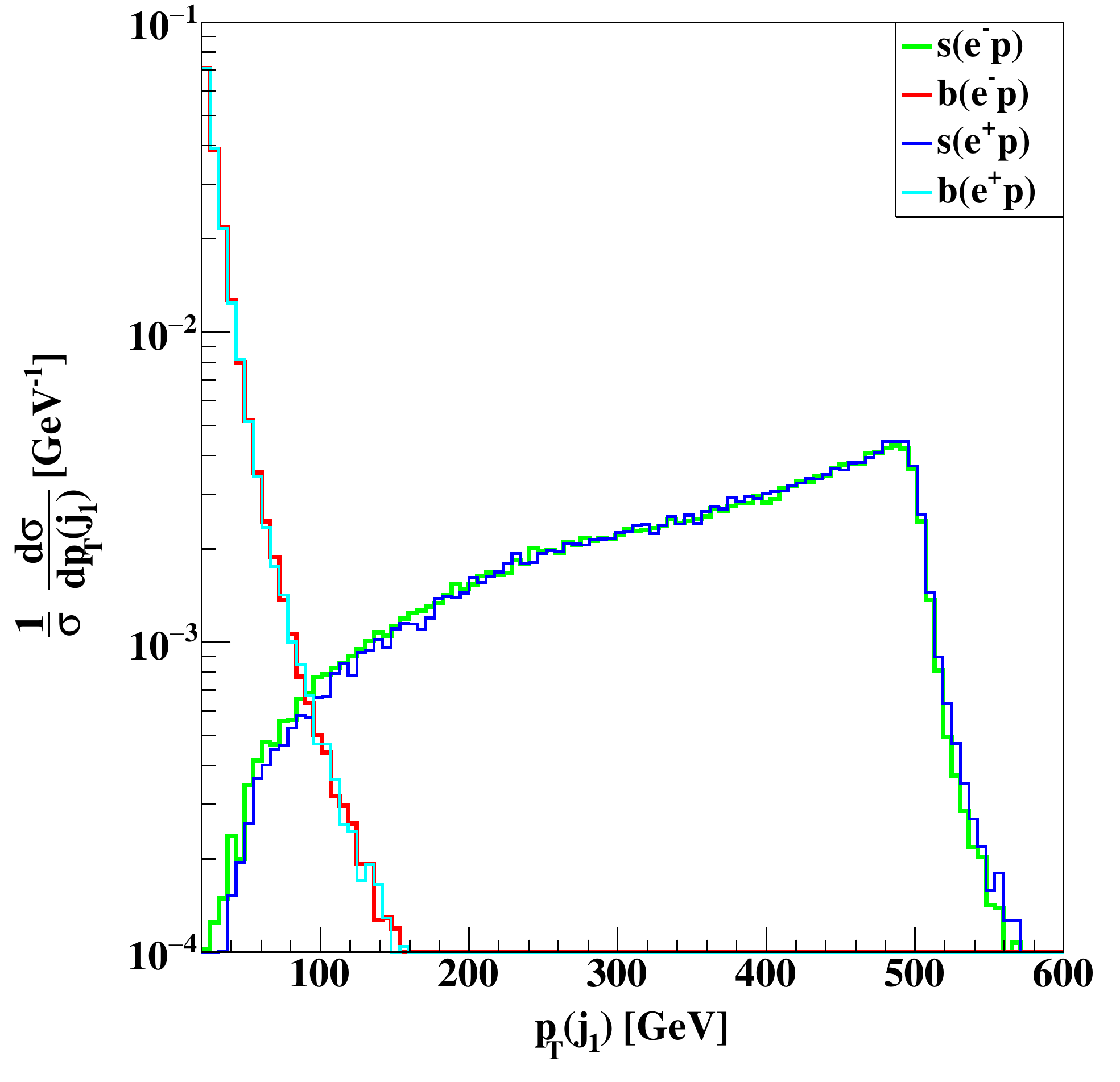}
 	\includegraphics[width=0.45\textwidth,height=0.3\textheight]{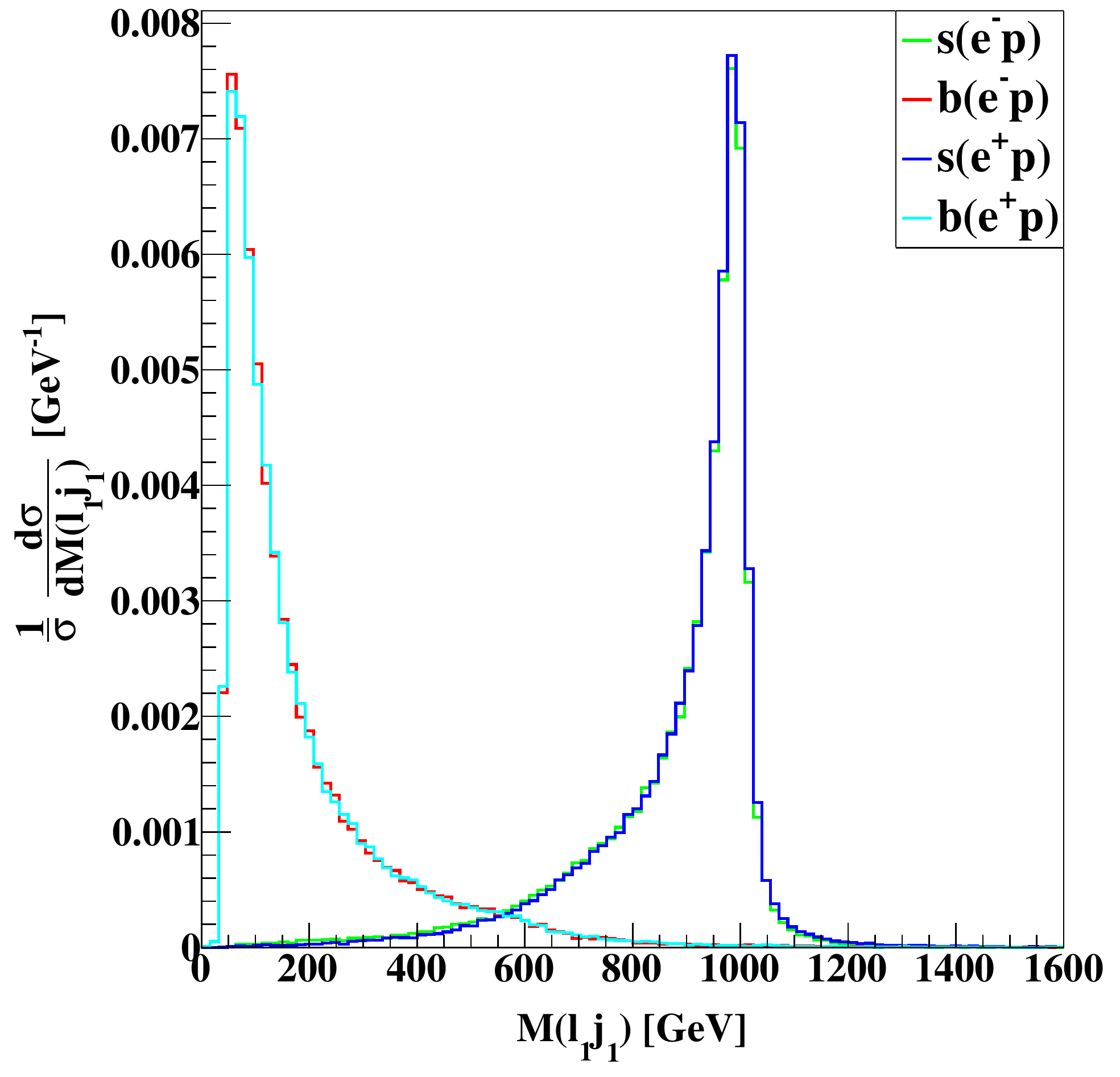}
 	\caption{\small{Normalised distribution of transverse momentum of leading lepton $p_T(l_1)$, leading jet $p_T(j_1)$ , invariant mass of leading lepton and jet $M(l_1j_1)$ for both signal and background with c.m.energy $\sqrt{s} = 3.46$ TeV.  Here the parameter set given in Table.~\ref{table:BP1} is considered. }}
 	\label{fig:distribution_fcc}
 \end{figure}
 
 Figs.~\ref{fig:distribution_lhec} and \ref{fig:distribution_fcc}  show the  normalised distributions of different kinematical variables~(transverse momentum of the leading lepton, leading jet, as well as the invariant mass distribution of the leading jet and leading lepton)~for LHeC  and FCC-eh, respectively. We have shown these distributions for $\text{BP}_1$. For LHeC and FCC-eh,  the distributions of background sample are similar for both the $e^-$ and $e^+$ case. For LHeC, the distributions of signal sample for $e^-$ and $e^+$ beam are different. For $e^-$ beam there are two peaks,  one lies in the lower $p_T$ region and the other in the higher  $p_T$ region. The peak lying in the lower $p_T$ region corresponds to the off-shell production of LQ.  For $e^+$ beam there is no second peak which implies that the off-shell production of LQ is less than that in case of $e^-$ beam. This occurs since  for $e^-p~(e^+p)$ collider, LQ couples to $e^--\bar{d}$ ($e^+-d$).  The $d$ quark carries greater fraction of proton momentum. From  Fig.~\ref{fig:distribution_fcc} it is evident that for  FCC-eh there are no such differences between the use of $e^-$ and $e^+$ beams due to availability of enough c.m. energy for LQ mass 1~TeV.
 \begin{table}[h]
 	\scalebox{0.9}{	\begin{tabular}{|c|c|c|c|c|c|}
 			\hline                
 			& \multicolumn{2}{c|}{ $e^- p \to l^-  j$ } && \multicolumn{2}{c|}{$e^+ p \to l^+  j$ } \\ \hline
 			&$\sigma^{\text{sig}}$ \text{[fb]} & $\sigma^{\text{bkg}}$ \text{[fb]}  &&$\sigma^{\text{sig}}$ \text{[fb]} & $\sigma^{\text{bkg}}$ \text{[fb]}     \\ \hline
 			before cut & 4.016& 2180&& 39.23 & 1440    \\ \hline
 			$c_1 : N_j \geq1 \ + N_{\ell}\geq1$& 3.01& 1644&&        29.85& 1079     \\ \hline
 			$c_2 : c_1 + p_T(\ell_{1})\geq 400$ &0.365 & 13.98&&     11.77& 6.54     \\ \hline
 			$c_3 : c_2 + p_T(j_{1})\geq 400$ &0.275 & 9.51&&         8.92 & 4.48     \\ \hline
 			$c_4 : c_4 +|M_{LQ}-M_{lj}|\leq100$  & 0.25&  5.13 &&                8.3& 2.534     \\ \hline
 			Significance for $\mathcal{L}= 1 \ \text{fb}^{-1}$ &\multicolumn{2}{c|}{0.107}&&\multicolumn{2}{c|}{2.5} \\ \hline
 	\end{tabular}}
 	\caption{Signal ($e^\pm p \to l^\pm  j$) and background cross-sections after different selection cuts at LHeC for $\text{BP}_1$.} \label{table:cuts}
 \end{table}
 
 \begin{table}[h]
 	\scalebox{0.9}{ \begin{tabular}{|c|c|c|c|c|c|}
 			\hline                
 			& \multicolumn{2}{c|}{$e^- p \to l^-  j$ } && \multicolumn{2}{c|}{ $e^+ p \to l^+  j$} \\ \hline
 			&$\sigma^{\text{sig}}$ \text{[fb]} & $\sigma^{\text{bkg}}$ \text{[fb]}  &&$\sigma^{\text{sig}}$ \text{[fb]} & $\sigma^{\text{bkg}}$ \text{[fb]}     \\ \hline
 			before cut &   395.08 & 10900&&   1246.4   &  9597    \\ \hline
 			$c_1 : N_j \geq1 \ + N_{\ell}\geq1$ & 354.41&9836.93&& 1119.03&  8652.58 \\ \hline
 			$c_2 : c_1 + p_T(\ell_{1})\geq 400$ &   180 & 839.141 && 578.13& 611.459       \\ \hline
 			$c_3 : c_2 + p_T(j_{1})\geq 400$ & 129.97 &618.963     &&  417.26 &    441.812  \\ \hline
 			$c_4 : c_3 +|M_{LQ}-M_{\ell j}|\leq100$  & 119.9 & 141.112&&  383.59 & 90.279   \\ \hline
 			Significance for $\mathcal{L}= 1 \ \text{fb}^{-1}$ &\multicolumn{2}{c|}{7.42}&&\multicolumn{2}{c|}{17.6} \\ \hline
 	\end{tabular}}
 	\caption{Signal ($e^\pm p \to l^\pm  j$ ) and background cross-sections after different selection cuts at FCC-eh for $\text{BP}_1$.} \label{table:cutsfcc}	
 \end{table}

 In Table.~\ref{table:cuts}, we have shown the variation of signal and background cross-sections after applying different selection cuts one by one at LHeC for $\text{BP}_1$. Before applying any selection cuts, the  signal and background cross-sections for the channel $e^- p \to l^-  j$ ($e^+ p \to l^+  j$) are 4.0164 fb (39.23 fb)
 and  2180 fb (1440 fb)\footnote{Note that, for the SM background, at the event generations level we impose two additional cuts on transverse momentum of leading lepton and jet, which are $p_{T}(l_1) > 200$ GeV and $p_{T}(j_1) > 200$ GeV to have better statistics as the SM background cross-section is very large.}, respectively. Note that due to the initial and final state radiations, additional jets will be present in the final
states considered. For signal selection we demand atleast one lepton and one jet in the final state. Then we impose a cut on transverse momentum of leading jet and lepton i.e.  $p_T(j _{1})\geq 400$ GeV and $p_T(l_{1})\geq 400$ GeV. This significantly reduces  the background cross-section. Finally we select the signal events by demanding a window
 on lepton-jet invariant mass in between $M_{\text{LQ}} \pm100 $GeV. After imposing all cuts the signal cross-section for the channel $e^- p \to l^-  j$ ($e^+ p \to l^+  j$) is reduced to one sixteenth (one fifth) of its initial value. Similarly Table.~\ref{table:cutsfcc} represents the result for FCC-eh.  We calculate the statistical significance using the expression: $\mathcal{S}=\dfrac{s}{\sqrt{s+b}}$, where s and b are number of signal and background events after all cuts, respectively. 
 \section{Results}
\begin{figure}
 	\includegraphics[width=7cm,height=6cm]{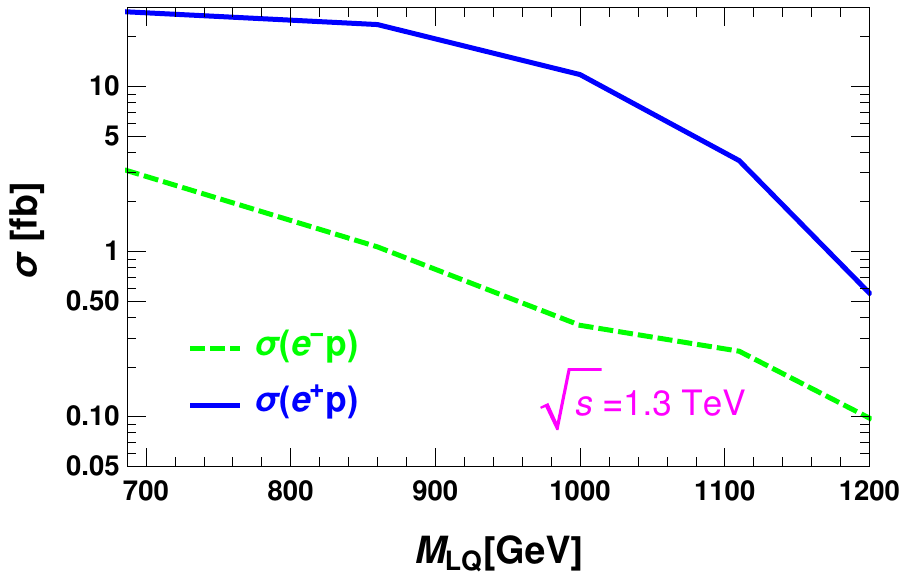}
 		\includegraphics[width=7cm,height=6cm]{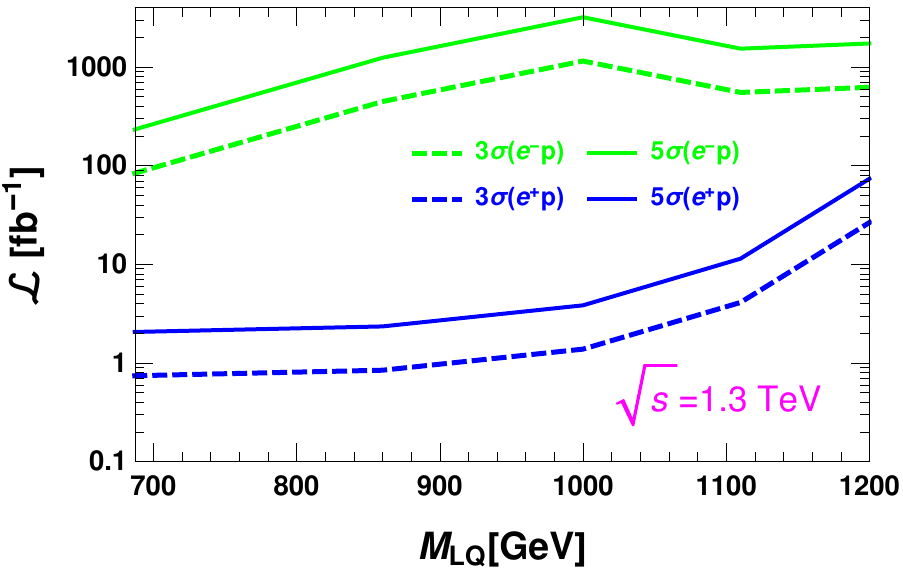}
 	\caption{Upper Panel: Signal cross-section after cut. Lower Panel: Required luminosity to reach $3\sigma$ and $5\sigma$ significance versus mass of the LQ with c.m.energy 1.3 TeV. Since at LHeC c.m.energy is low to get a better signal cross-section we use the optimum values of the couplings for a given LQ mass, which are given in Table.~\ref{table:mlq_y_z}.  } \label{fig:signf}
 \end{figure}

\begin{figure}
 		\includegraphics[width=7cm,height=6cm]{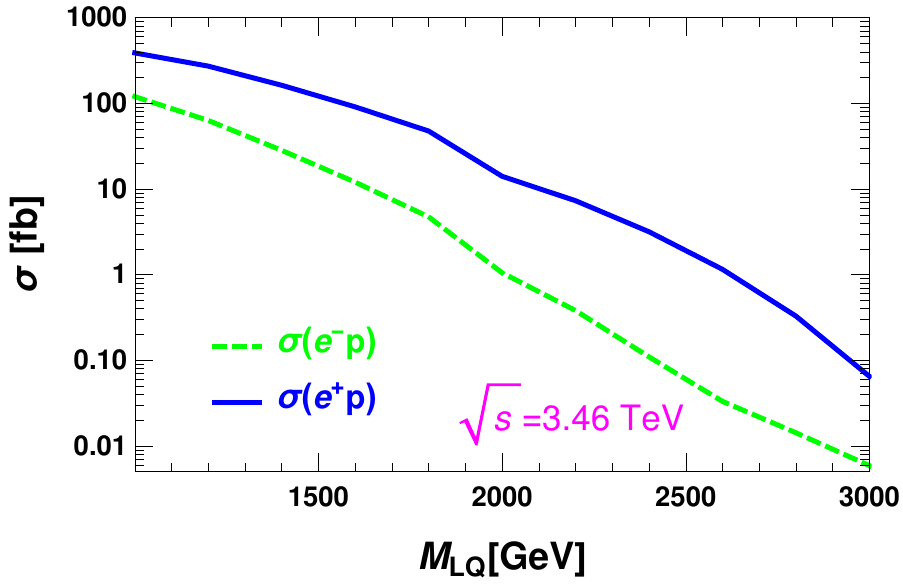}
 	\includegraphics[width=7cm,height=6cm]{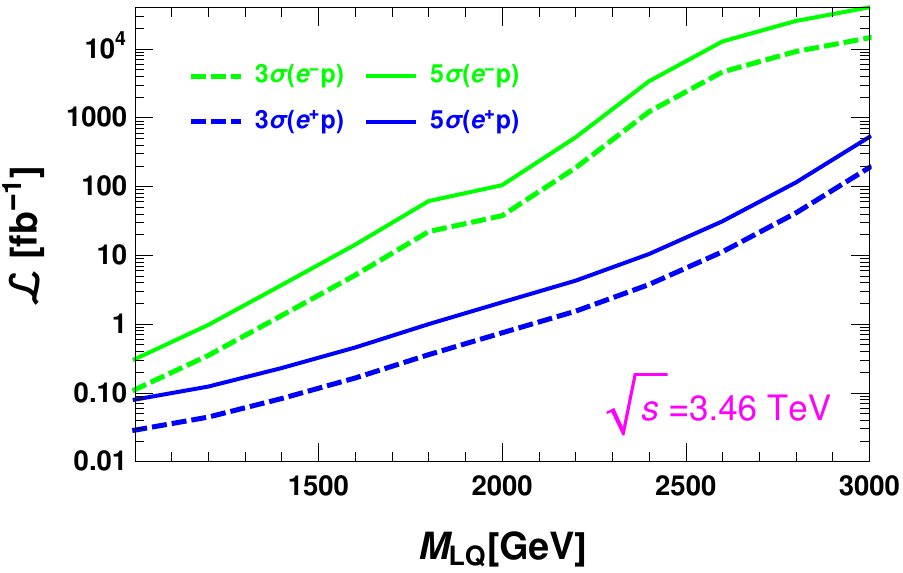}
 	\caption{Upper Panel: Signal cross-section after cut. Lower Panel: Required luminosity to reach $3\sigma$ and $5\sigma$ significance versus mass of the LQ with c.m.energy 3.46 TeV. In this case we consider the same couplings for all LQ masses as given in Table.~\ref{table:BP1} for $M_{\text{LQ}}$=1 TeV.  } \label{fig:signf_fcc}
 \end{figure}

 We discuss the discovery prospects of LQs at LHeC and FCC-eh in the mass range 700 GeV-1200 GeV and 1 TeV-3 TeV, respectively. In the upper panel of Figs.~\ref{fig:signf} and \ref{fig:signf_fcc}, we have shown the signal cross-section after all the cuts at LHeC and FCC-eh, respectively.
 In the lower panel of Fig.~\ref{fig:signf},  we have shown the required luminosity ($\mathcal{L}$) to achieve $3\sigma$ and $5\sigma$ significance versus $M_{\text{LQ}}$ for LHeC. Yukawa coupling  $Y_{11}$ has been fixed to its upper limit, $0.34\frac{M_{\text{LQ}}}{1\,\text{TeV}}$  for a given $M_{\text{LQ}}$ according to APV constraint. $Z_{11}$ has been fixed to its lower limit for given  $M_{\text{LQ}}$ and $Y_{11}$,  according to the  upper-bound from LHC search. For $M_{\text{LQ}}\geq 1$~TeV, we use the same selection cuts as given in Table.~\ref{table:cuts}. For $M_{\text{LQ}}<1$~TeV, we again apply the same set of cuts except the cuts  $c_2$ and $c_3$. Here  cuts $c_2$ and $c_3$ are defined as $c_1 + p_T(\ell_{1})\geq 300$ GeV and  $c_2 + p_T(j_{1})\geq 300$ GeV, respectively.
 
 Similarly, Fig.~\ref{fig:signf_fcc} shows the variation of  required luminosity ($\mathcal{L}$) to achieve $3\sigma$ and $5\sigma$ significance, as a function of $M_{\text{LQ}}$ for FCC-eh. For $M_{\text{LQ}}<2$ TeV, we use same selection cuts as given in Table.~\ref{table:cutsfcc}.  For  $M_{\text{LQ}}\ge2$ TeV, we define $c_2$ and $c_3$ as  $c_1 + p_T(\ell_{1})\geq 800$ GeV and  $c_2 + p_T(j_{1})\geq 800$ GeV, respectively.
 
 We find that, this final state has reasonably good discovery prospect even after giving a generic set of cuts. For LHeC with $e^-$ beam, its difficult to probe LQ due to small cross-section. From the lower panel of Fig.~\ref{fig:signf}, it is evident that to probe higher LQ masses with $5\sigma$ significance, the required luminosity is very high in spite of smaller SM background, as the signal cross section
is itself small for higher LQ masses. On the contrary, for LHeC with $e^+$ beam, due to larger signal cross-sections, it is possible to probe the LQ mass upto 1.2 TeV at more than $5\sigma$ significance with integrated luminosity less than $100\,\text{fb}^{-1}$. 

For FCC-eh, even with $e^-$ beam, we can probe LQ mass upto 2.3 TeV at more than $5\sigma$ significance with integrated luminosity less than $1000\,\text{fb}^{-1}$. For higher LQ masses, the required luminosity to probe with $5\sigma$ significance increases quickly. For the case of $e^+$ beam, due to enhanced cross-section we can very easily probe the LQ mass upto 3 TeV at more than $5\sigma$ significance with integrated luminosity less than $500\,\text{fb}^{-1}$. Note that, the production cross-section of $\tilde{R}_2$ at the $ep$ colliders and its decay vary with respect to the Yukawa couplings ($Y_{11}$ and $Z_{11}$). Therefore, the discovery reach of LHeC and FCC-eh presented in this analysis depend on the chosen values of Yukawa couplings. Several studies have been carried out to predict the discovery prospect of other types of LQs at HL-LHC. The studies~\cite{Chandak:2019iwj,Allanach:2019zfr} shows that, scalar LQs of masses upto $2$ TeV can be discovered at HL-LHC.

Finally, we would like to point out that as with the polarized electron or positron beams, $lj$ production cross-sections are enhanced by almost a factor of two, we can probe even higher LQ masses compared to the case of unpolarized beams. We evaluate the asymmetry between the production cross-section of LQ at $e^+p$ and $e^-p$ colliders, which is defined as $\mathcal{A}_{ep}=\frac{\sigma(e^+p)-\sigma(e^-p)}{\sigma(e^+p)+\sigma(e^-p)}$. It is found to be positive (which is evident from the upper panel of Figs.~\ref{fig:signf} and \ref{fig:signf_fcc}), consistent with the fermion number of $\tilde{R}_2$  LQ . 

\section{conclusions}
\label{conclusion}
In this work we study the discovery prospect of $\tilde{R}_2$ class of LQ models at the proposed $ep$ colliders such as LHeC and FCC-eh. This type of LQ can be copiously produced at $ep$ colliders, due to its interaction with the electron and down type quarks. There are many possible final states for this type of LQ model. We specifically focus on LQ production and its decay to a lepton and a jet. For this final state, we expect higher statistical significance at an $ep$ collider compared to $pp$ collider due to higher production cross-section as well as lower SM backgrounds. We find that at LHeC with $e^+$ beam, we can probe the LQ mass upto 1.2 TeV at
more than $5\sigma$ significance. For FCC-eh, with $e^-$ beam we can probe LQ mass upto 3 TeV with $5\sigma$ significance, but the required luminosity is large. On the contrary, at FCC-eh with $e^+$ beam, we can easily probe the LQ mass upto 3 TeV at more than $5\sigma$ significance with nominal integrated luminosity. Note that at an $ep$ collider, polarization of the electron or positron beams
can result in a substantial increase in the LQ production and the resulting
lepton-jet production cross-sections.

\bibliographystyle{utphys}
\bibliography{bibitem.bib}

\providecommand{\href}[2]{#2}\begingroup\raggedright\begin{thebibliography}{10}

\bibitem{Pati:1973uk}
J.~C. Pati and A.~Salam, ``{Unified Lepton-Hadron Symmetry and a Gauge Theory
  of the Basic Interactions},''
\href{http://dx.doi.org/10.1103/PhysRevD.8.1240}{{\em Phys. Rev.} {\bfseries
  D8} (1973) 1240--1251}.

\bibitem{Georgi:1974sy}
H.~Georgi and S.~L. Glashow, ``{Unity of All Elementary Particle Forces},''
\href{http://dx.doi.org/10.1103/PhysRevLett.32.438}{{\em Phys. Rev. Lett.}
  {\bfseries 32} (1974) 438--441}.

\bibitem{Fritzsch:1974nn}
H.~Fritzsch and P.~Minkowski, ``{Unified Interactions of Leptons and
  Hadrons},''
\href{http://dx.doi.org/10.1016/0003-4916(75)90211-0}{{\em Annals Phys.}
  {\bfseries 93} (1975) 193--266}.

\bibitem{Mohapatra:1979nn}
R.~N. Mohapatra and B.~Sakita, ``{SO(2n) Grand Unification in an SU(N)
  Basis},'' \href{http://dx.doi.org/10.1103/PhysRevD.21.1062}{{\em Phys. Rev.}
  {\bfseries D21} (1980) 1062}.
[,283(1979)].

\bibitem{Wilczek:1981iz}
F.~Wilczek and A.~Zee, ``{Families from Spinors},''
\href{http://dx.doi.org/10.1103/PhysRevD.25.553}{{\em Phys. Rev.} {\bfseries
  D25} (1982) 553}.

\bibitem{Dimopoulos:1979es}
S.~Dimopoulos and L.~Susskind, ``{Mass Without Scalars},''
  \href{http://dx.doi.org/10.1016/0550-3213(79)90364-X}{{\em Nucl. Phys.}
  {\bfseries B155} (1979) 237--252}.
[2,930(1979)].

\bibitem{Dimopoulos:1979sp}
S.~Dimopoulos, ``{Technicolored Signatures},''
\href{http://dx.doi.org/10.1016/0550-3213(80)90277-1}{{\em Nucl. Phys.}
  {\bfseries B168} (1980) 69--92}.

\bibitem{Farhi:1980xs}
E.~Farhi and L.~Susskind, ``{Technicolor},''
\href{http://dx.doi.org/10.1016/0370-1573(81)90173-3}{{\em Phys. Rept.}
  {\bfseries 74} (1981) 277}.

\bibitem{Georgi:1981xw}
H.~Georgi and S.~L. Glashow, ``{Unextended Technicolor and Unification},''
\href{http://dx.doi.org/10.1103/PhysRevLett.47.1511}{{\em Phys. Rev. Lett.}
  {\bfseries 47} (1981) 1511}.

\bibitem{Dorsner:2016wpm}
I.~Doršner, S.~Fajfer, A.~Greljo, J.~F. Kamenik, and N.~Košnik, ``{Physics of
  leptoquarks in precision experiments and at particle colliders},''
  \href{http://dx.doi.org/10.1016/j.physrep.2016.06.001}{{\em Phys. Rept.}
  {\bfseries 641} (2016) 1--68},
\href{http://arxiv.org/abs/1603.04993}{{\ttfamily arXiv:1603.04993 [hep-ph]}}.

\bibitem{Arnold:2012sd}
J.~M. Arnold, B.~Fornal, and M.~B. Wise, ``{Simplified models with baryon
  number violation but no proton decay},''
  \href{http://dx.doi.org/10.1103/PhysRevD.87.075004}{{\em Phys. Rev.}
  {\bfseries D87} (2013) 075004},
\href{http://arxiv.org/abs/1212.4556}{{\ttfamily arXiv:1212.4556 [hep-ph]}}.

\bibitem{Cox:2016epl}
P.~Cox, A.~Kusenko, O.~Sumensari, and T.~T. Yanagida, ``{SU(5) Unification with
  TeV-scale Leptoquarks},''
  \href{http://dx.doi.org/10.1007/JHEP03(2017)035}{{\em JHEP} {\bfseries 03}
  (2017) 035},
\href{http://arxiv.org/abs/1612.03923}{{\ttfamily arXiv:1612.03923 [hep-ph]}}.

\bibitem{Raj:2016aky}
N.~Raj, ``{Anticipating nonresonant new physics in dilepton angular spectra at
  the LHC},'' \href{http://dx.doi.org/10.1103/PhysRevD.95.015011}{{\em Phys.
  Rev.} {\bfseries D95} no.~1, (2017) 015011},
\href{http://arxiv.org/abs/1610.03795}{{\ttfamily arXiv:1610.03795 [hep-ph]}}.

\bibitem{Chandak:2019iwj}
K.~Chandak, T.~Mandal, and S.~Mitra, ``{Hunting for scalar leptoquarks with
  boosted tops and light leptons},''
  \href{http://dx.doi.org/10.1103/PhysRevD.100.075019}{{\em Phys. Rev.}
  {\bfseries D100} no.~7, (2019) 075019},
\href{http://arxiv.org/abs/1907.11194}{{\ttfamily arXiv:1907.11194 [hep-ph]}}.

\bibitem{Vignaroli:2018lpq}
N.~Vignaroli, ``{Seeking leptoquarks in the $\bf t\bar{t}$ plus missing energy
  channel at the high-luminosity LHC},''
  \href{http://dx.doi.org/10.1103/PhysRevD.99.035021}{{\em Phys. Rev.}
  {\bfseries D99} no.~3, (2019) 035021},
\href{http://arxiv.org/abs/1808.10309}{{\ttfamily arXiv:1808.10309 [hep-ph]}}.

\bibitem{Allanach:2019zfr}
B.~C. Allanach, T.~Corbett, and M.~Madigan, ``{Sensitivity of Future Hadron
  Colliders to Leptoquark Pair Production in the Di-Muon Di-Jets Channel},''
  \href{http://dx.doi.org/10.1140/epjc/s10052-020-7722-3}{{\em Eur. Phys. J.}
  {\bfseries C80} no.~2, (2020) 170},
\href{http://arxiv.org/abs/1911.04455}{{\ttfamily arXiv:1911.04455 [hep-ph]}}.

\bibitem{Mandal:2018kau}
T.~Mandal, S.~Mitra, and S.~Raz, ``{$R_{D^{(*)}}$ motivated $\mathcal{S}_1$
  leptoquark scenarios: Impact of interference on the exclusion limits from LHC
  data},'' \href{http://dx.doi.org/10.1103/PhysRevD.99.055028}{{\em Phys. Rev.}
  {\bfseries D99} no.~5, (2019) 055028},
\href{http://arxiv.org/abs/1811.03561}{{\ttfamily arXiv:1811.03561 [hep-ph]}}.

\bibitem{Cerri:2018ypt}
A.~Cerri {\em et~al.}, ``{Report from Working Group 4},''
  \href{http://dx.doi.org/10.23731/CYRM-2019-007.867}{{\em CERN Yellow Rep.
  Monogr.} {\bfseries 7} (2019) 867--1158},
\href{http://arxiv.org/abs/1812.07638}{{\ttfamily arXiv:1812.07638 [hep-ph]}}.

\bibitem{AbelleiraFernandez:2012cc}
{\bfseries LHeC Study Group} Collaboration, J.~L. Abelleira~Fernandez {\em
  et~al.}, ``{A Large Hadron Electron Collider at CERN: Report on the Physics
  and Design Concepts for Machine and Detector},''
  \href{http://dx.doi.org/10.1088/0954-3899/39/7/075001}{{\em J. Phys.}
  {\bfseries G39} (2012) 075001},
\href{http://arxiv.org/abs/1206.2913}{{\ttfamily arXiv:1206.2913
  [physics.acc-ph]}}.

\bibitem{Azuelos:2018syu}
G.~Azuelos, M.~D'Onofrio, O.~Fischer, and J.~Zurita, ``{BSM physics at the LHeC
  and the FCC-eh},'' \href{http://dx.doi.org/10.22323/1.316.0190}{{\em PoS}
  {\bfseries DIS2018} (2018) 190},
\href{http://arxiv.org/abs/1807.01618}{{\ttfamily arXiv:1807.01618 [hep-ph]}}.

\bibitem{Zimmermann:2014qxa}
F.~Zimmermann, M.~Benedikt, D.~Schulte, and J.~Wenninger,
  \href{http://dx.doi.org/10.18429/JACoW-IPAC2014-MOXAA01}{``{Challenges for
  Highest Energy Circular Colliders},''} in {\em {Proceedings, 5th
  International Particle Accelerator Conference (IPAC 2014): Dresden, Germany,
  June 15-20, 2014}}, p.~MOXAA01.
\newblock 2014.
\newblock
\url{http://jacow.org/IPAC2014/papers/moxaa01.pdf}.
\newblock

\bibitem{LHeCFCCehBSM}
 \url{https://twiki.cern.ch/twiki/bin/viewauth/LHeC/LHeCFCCehBSM}.

\bibitem{Buchmuller:1986zs}
W.~Buchmuller, R.~Ruckl, and D.~Wyler, ``{Leptoquarks in Lepton - Quark
  Collisions},'' \href{http://dx.doi.org/10.1016/S0370-2693(99)00014-3,
  10.1016/0370-2693(87)90637-X}{{\em Phys. Lett.} {\bfseries B191} (1987)
  442--448}.
[Erratum: Phys. Lett.B448,320(1999)].

\bibitem{Buchmuller:1986iq}
W.~Buchmuller and D.~Wyler, ``{Constraints on SU(5) Type Leptoquarks},''
\href{http://dx.doi.org/10.1016/0370-2693(86)90771-9}{{\em Phys. Lett.}
  {\bfseries B177} (1986) 377--382}.

\bibitem{Becirevic:2016yqi}
D.~Bečirević, S.~Fajfer, N.~Košnik, and O.~Sumensari, ``{Leptoquark model to
  explain the $B$-physics anomalies, $R_K$ and $R_D$},''
  \href{http://dx.doi.org/10.1103/PhysRevD.94.115021}{{\em Phys. Rev.}
  {\bfseries D94} no.~11, (2016) 115021},
\href{http://arxiv.org/abs/1608.08501}{{\ttfamily arXiv:1608.08501 [hep-ph]}}.

\bibitem{Dorsner:2014axa}
I.~Dorsner, S.~Fajfer, and A.~Greljo, ``{Cornering Scalar Leptoquarks at
  LHC},'' \href{http://dx.doi.org/10.1007/JHEP10(2014)154}{{\em JHEP}
  {\bfseries 10} (2014) 154},
\href{http://arxiv.org/abs/1406.4831}{{\ttfamily arXiv:1406.4831 [hep-ph]}}.

\bibitem{Babu:2019mfe}
K.~S. Babu, P.~S.~B. Dev, S.~Jana, and A.~Thapa, ``{Non-Standard Interactions
  in Radiative Neutrino Mass Models},''
\href{http://arxiv.org/abs/1907.09498}{{\ttfamily arXiv:1907.09498 [hep-ph]}}.

\bibitem{Dorsner:2011ai}
I.~Dorsner, J.~Drobnak, S.~Fajfer, J.~F. Kamenik, and N.~Kosnik, ``{Limits on
  scalar leptoquark interactions and consequences for GUTs},''
  \href{http://dx.doi.org/10.1007/JHEP11(2011)002}{{\em JHEP} {\bfseries 11}
  (2011) 002},
\href{http://arxiv.org/abs/1107.5393}{{\ttfamily arXiv:1107.5393 [hep-ph]}}.

\bibitem{Sirunyan:2017yrk}
{\bfseries CMS} Collaboration, A.~M. Sirunyan {\em et~al.}, ``{Search for
  third-generation scalar leptoquarks and heavy right-handed neutrinos in final
  states with two tau leptons and two jets in proton-proton collisions at $
  \sqrt{s}=13 $ TeV},'' \href{http://dx.doi.org/10.1007/JHEP07(2017)121}{{\em
  JHEP} {\bfseries 07} (2017) 121},
\href{http://arxiv.org/abs/1703.03995}{{\ttfamily arXiv:1703.03995 [hep-ex]}}.

\bibitem{Aaboud:2016qeg}
{\bfseries ATLAS} Collaboration, M.~Aaboud {\em et~al.}, ``{Search for scalar
  leptoquarks in pp collisions at $\sqrt{s}$ = 13 TeV with the ATLAS
  experiment},'' \href{http://dx.doi.org/10.1088/1367-2630/18/9/093016}{{\em
  New J. Phys.} {\bfseries 18} no.~9, (2016) 093016},
\href{http://arxiv.org/abs/1605.06035}{{\ttfamily arXiv:1605.06035 [hep-ex]}}.

\bibitem{Sirunyan:2018kzh}
{\bfseries CMS} Collaboration, A.~M. Sirunyan {\em et~al.}, ``{Constraints on
  models of scalar and vector leptoquarks decaying to a quark and a neutrino at
  $\sqrt{s}=$ 13 TeV},''
  \href{http://dx.doi.org/10.1103/PhysRevD.98.032005}{{\em Phys. Rev.}
  {\bfseries D98} no.~3, (2018) 032005},
\href{http://arxiv.org/abs/1805.10228}{{\ttfamily arXiv:1805.10228 [hep-ex]}}.

\bibitem{Sirunyan:2018jdk}
{\bfseries CMS} Collaboration, A.~M. Sirunyan {\em et~al.}, ``{Search for a
  singly produced third-generation scalar leptoquark decaying to a $\tau$
  lepton and a bottom quark in proton-proton collisions at $\sqrt{s} =$ 13
  TeV},'' \href{http://dx.doi.org/10.1007/JHEP07(2018)115}{{\em JHEP}
  {\bfseries 07} (2018) 115},
\href{http://arxiv.org/abs/1806.03472}{{\ttfamily arXiv:1806.03472 [hep-ex]}}.

\bibitem{ATLAS:2017vke}
{\bfseries CMS, ATLAS} Collaboration, F.~Romeo, ``{Searches for new physics in
  lepton plus jet final states in ATLAS and CMS},'' in {\em {2017 International
  Workshop on Baryon and Lepton Number Violation: From the Cosmos to the LHC
  (BLV 2017) Cleveland, Ohio, USA, May 15-18, 2017}}.
\newblock 2017.
\newblock
\href{http://arxiv.org/abs/1709.00229}{{\ttfamily arXiv:1709.00229 [hep-ex]}}.
\newblock

\bibitem{Sirunyan:2018btu}
{\bfseries CMS} Collaboration, A.~M. Sirunyan {\em et~al.}, ``{Search for pair
  production of first-generation scalar leptoquarks at $\sqrt{s} =$ 13 TeV},''
  \href{http://dx.doi.org/10.1103/PhysRevD.99.052002}{{\em Phys. Rev.}
  {\bfseries D99} no.~5, (2019) 052002},
\href{http://arxiv.org/abs/1811.01197}{{\ttfamily arXiv:1811.01197 [hep-ex]}}.

\bibitem{Alloul:2013bka}
A.~Alloul, N.~D. Christensen, C.~Degrande, C.~Duhr, and B.~Fuks, ``{FeynRules
  2.0 - A complete toolbox for tree-level phenomenology},''
  \href{http://dx.doi.org/10.1016/j.cpc.2014.04.012}{{\em Comput. Phys.
  Commun.} {\bfseries 185} (2014) 2250--2300},
\href{http://arxiv.org/abs/1310.1921}{{\ttfamily arXiv:1310.1921 [hep-ph]}}.

\bibitem{Alwall:2014hca}
J.~Alwall, R.~Frederix, S.~Frixione, V.~Hirschi, F.~Maltoni, O.~Mattelaer,
  H.~S. Shao, T.~Stelzer, P.~Torrielli, and M.~Zaro, ``{The automated
  computation of tree-level and next-to-leading order differential cross
  sections, and their matching to parton shower simulations},''
  \href{http://dx.doi.org/10.1007/JHEP07(2014)079}{{\em JHEP} {\bfseries 07}
  (2014) 079},
\href{http://arxiv.org/abs/1405.0301}{{\ttfamily arXiv:1405.0301 [hep-ph]}}.

\bibitem{Sjostrand:2006za}
T.~Sjostrand, S.~Mrenna, and P.~Z. Skands, ``{PYTHIA 6.4 Physics and Manual},''
  \href{http://dx.doi.org/10.1088/1126-6708/2006/05/026}{{\em JHEP} {\bfseries
  05} (2006) 026},
\href{http://arxiv.org/abs/hep-ph/0603175}{{\ttfamily arXiv:hep-ph/0603175
  [hep-ph]}}.

\bibitem{deFavereau:2013fsa}
{\bfseries DELPHES 3} Collaboration, J.~de~Favereau, C.~Delaere, P.~Demin,
  A.~Giammanco, V.~Lemaître, A.~Mertens, and M.~Selvaggi, ``{DELPHES 3, A
  modular framework for fast simulation of a generic collider experiment},''
  \href{http://dx.doi.org/10.1007/JHEP02(2014)057}{{\em JHEP} {\bfseries 02}
  (2014) 057},
\href{http://arxiv.org/abs/1307.6346}{{\ttfamily arXiv:1307.6346 [hep-ex]}}.

\bibitem{Brun:1997pa}
R.~Brun and F.~Rademakers, ``{ROOT: An object oriented data analysis
  framework},''
\href{http://dx.doi.org/10.1016/S0168-9002(97)00048-X}{{\em Nucl. Instrum.
  Meth.} {\bfseries A389} (1997) 81--86}.

\end{thebibliography}\endgroup

\end{document}